\documentclass[superscriptaddress,column,showpacs,a4paper,
amssymb,amsmath,nobibnotes,aps,prd,longbibliography,
showkeys,
nofootinbib,notitlepage]{revtex4-1}
\usepackage{bigints}
\usepackage{graphicx}
\usepackage{float}
\usepackage{epsf}
\usepackage{bm}
\usepackage{amsmath}
\usepackage{amsfonts}
\usepackage{amssymb}
\usepackage{epstopdf}
\usepackage{natbib}
\usepackage{color}
\usepackage{placeins}
\usepackage{amsmath}
\numberwithin{equation}{section}

\providecommand{\U}[1]{\protect\rule{.1in}{.1in}}
\newcommand{\be}{\begin{equation}}
\newcommand{\ee}{\end{equation}}

\newcommand{\mincir}{\raise
-3.truept\hbox{\rlap{\hbox{$\sim$}}\raise4.truept\hbox{$<$}\ }}
\newcommand{\magcir}{\raise
-3.truept\hbox{\rlap{\hbox{$\sim$}}\raise4.truept\hbox{$>$}\ }}

\newtheorem{remark}{Remark}[section]
\usepackage{xcolor}
\usepackage{hyperref}
\hypersetup{colorlinks=true, linkcolor=blue,citecolor=magenta}

\begin{document}


\title{{A Bottom-Up Derivation of Einstein's Field Equations in Harmonic Gauge: \\ From Wave Dynamics to Pseudo-Riemannian Geometry}}

\author{Jaume de Haro}
\email{jaime.haro@upc.edu}
\affiliation{Departament de Matem\`atiques, Universitat Polit\`ecnica de Catalunya, 
Carrer de Colom 11, 08222 Terrassa, Barcelona, Spain.}

\begin{abstract}
We present a constructive, bottom-up reconstruction of General Relativity, deriving Einstein's field equations in harmonic coordinates from dynamical consistency and the non-linear self-interaction of the gravitational field. 
Rather than postulating Pseudo-Riemannian geometry \textit{a priori}, we begin with the proper-time formulation of free fall to derive the linearized wave equation for perturbations of the Minkowski invariant, subsequently extending it to the full non-linear regime. By replacing the d'Alembertian with the Laplace--Beltrami operator and enforcing local energy-momentum conservation in arbitrary coordinates, 
we show that energy conservation forces an overconstrained algebraic system on the five-dimensional vector space of irreducible non-tensorial quadratic first-derivative combinations $U^{(k)}_{\mu\nu}$. This system admits a unique solution with coefficients $(c_1,c_2,c_3,c_4,c_5) = (1, 1, 1/2, 1, -1)$, through which the full Ricci tensor and the trace-reversed Einstein field equations emerge as the unique minimal second-order completion within the specified quadratic first-derivative ansatz and harmonic gauge. We provide a historical critique of Einstein's path in the Zürich Notebook (1912--1913) and his subsequent 1915 reliance on gravitational pseudo-tensors, and compare our constructive result with the spin-$2$ quantum field theoretic program of Feynman and Weinberg.
Our results demonstrate that spacetime curvature need not be regarded as a foundational a \textit{priori} postulate of gravitation, but can be rigorously understood as the natural macroscopic geometrical representation of dynamical consistency.
\end{abstract}

\vspace{0.5cm}


\maketitle

 \section{Introduction}
The standard presentation of General Relativity typically begins with a profound epistemological leap: space and time are fused into a Pseudo-Riemannian manifold (to be precise, a Lorentz manifold), and gravity is postulated from the outset as the manifestation of spacetime curvature \cite{MTW1973,Synge1960,Hawking1973,Sachs1977,Wald1984,Carroll2019}. While geometrically elegant, this top-down approach obscures the underlying physical mechanisms that compel the gravitational field to couple non-linearly to itself, creating a conceptual divide between gravitation and standard continuum field theories \cite{Feynman1995,Deser1970,Weinberg1965,Weinberg1972,Gupta1954,Kraichnan1955}. Historically, Albert Einstein did not arrive at his field equations through pure differential geometry; rather, his path was a tortuous struggle between physical intuition, weak-field limits, coordinate constraints, and energy conservation laws \cite{Girard1981,Norton1984,Renn2007,Straumann2011,Norton2020}.

In this work, we reverse the standard paradigm. We show that the core of General Relativity can be systematically constructed from a bottom-up physical program centered on proper time, wave propagation, and dynamical consistency \cite{Haro2026a}. By starting from Galilean free fall and assuming that test particles follow their \textit{conatus}---that is, their intrinsic physical tendency to preserve their existence, which dynamically translates into pursuing worldlines along which the particle's own clock locally maximizes its accumulated ticks---we establish how static mass distributions alter the proper-time structure. This operational formulation allows us to derive, within the weak-field approximation, the exact invariant ($ds^2$) for a reference frame at rest under the influence of a static gravitational field without postulating geometric axioms \textit{a priori}.

The  invariant corresponding to moving source distributions is subsequently obtained by applying kinematic Lorentz transformations to the elementary static perturbations. Because these boosts describe physical changes of frame on the background Minkowski spacetime, the resulting linear potentials naturally satisfy a linear wave equation sourced by the matter stress-energy tensor defined in flat space. 
By imposing the conservation of this stress-energy tensor, it follows directly that the linearized invariant satisfies the linear harmonic condition, historically also designated as the Hilbert gauge condition \cite{Einstein1916b, Hilbert1917}.

The harmonic gauge, 
far from being a mere coordinate convenience, holds a distinguished status in both the mathematical foundation and the practical application of General Relativity.
 It was precisely by casting the field equations in harmonic coordinates that Yvonne Choquet-Bruhat, building on earlier work on the Cauchy problem by André Lichnerowicz and others \cite{Lichnerowicz1952}, achieved the landmark proof of local existence and geometric uniqueness—up to diffeomorphisms—of vacuum solutions, establishing the well-posedness of the Cauchy problem for Einstein's theory \cite{ChoquetBruhat2009}.
In modern gravitational physics, this gauge remains indispensable for describing wave propagation and radiation; it underpins the post-Minkowskian and post-Newtonian formalisms developed by Luc Blanchet, Thibault Damour, and Clifford Will, enabling the high-precision extraction of gravitational wave templates from compact binary coalescences \cite{Blanchet2014,Will,Poisson,Damour,Deruelle}.
Converging naturally with this physical and mathematical legacy, our constructive approach does not impose the harmonic gauge \emph{ad hoc}; rather, it extends the linearized harmonic condition already forced by energy-momentum conservation to its exact non-linear form as the essential structural framework for wave propagation.

Once the linear wave equations are established, the central challenge is to construct the full non-linear equations encoding gravitational self-interaction---the fundamental requirement that the energy density of the gravitational field must itself act as a gravitational source. As Einstein pointed out in 1911 \cite{Einstein1911}, gravity alters the flow of proper time and consequently modifies the flat Minkowski invariant $ds_{\rm Min}^2$ of Special Relativity \cite{Minkowski1909}. Because the standard d'Alembertian governs wave propagation strictly with respect to $ds_{\rm Min}^2$, the gravitational modification of the proper-time invariant $ds^2 = g_{\mu\nu}dx^\mu dx^\nu$---where, aligning with his line of thought, $g_{\mu\nu}$ can be understood as the gravitational potentials---demands a consistent generalization of the wave operator itself, which is naturally provided by the Laplace--Beltrami operator \cite{Beltrami1868}.
 Casting the non-linear wave dynamics in the harmonic gauge \cite{deDonder1921} through the same wave-operator structure employed in \cite{Thorne}, we couple this operator to the complete 5-dimensional vector space of independent irreducible quadratic combinations $U^{(k)}$ built from first-order derivatives of the gravitational potentials $g_{\mu\nu}$, whose algebraic structure and completeness are strictly governed by Weyl's fundamental theorems of invariant theory \cite{Weyl1939}.

By enforcing the exact local conservation of matter energy-momentum, $\nabla_\mu T^{\mu}_{\nu} = 0$, as an inescapable consistency condition---where $\nabla_\mu$ acts as a purely dynamical differential operator reflecting local energy-momentum balance and free-fall particle dynamics, as derived in Subsection~\ref{subsection:A}, and which, in geometrical terms, is precisely the well-known Levi-Civita connection \cite{RicciLeviCivita1900}---we prove that the quadratic self-interaction blocks $U^{(k)}$ assemble uniquely into the exact second-order differential operator identified with the Ricci tensor $R_{\mu\nu}$. This procedure profoundly mirrors Einstein's methodological ambition in his 1913 \textit{Entwurf} theory \cite{Grossmann1914}---seeking a physically motivated field operator proportional to the (trace-reversed) stress-energy tensor---but succeeds precisely by allowing the dynamical balance law to dictate the unique algebraic combination of the self-interaction terms.

Taken together, these steps articulate a relational conception of gravitation that guides the entire program of this paper. Rather than treating gravity as an intrinsic property possessed by an isolated body or as a kinematic constraint imposed by an a priori curved manifold, we frame it through two complementary operational notions: the \emph{conatus} of a physical system---its active tendency to persevere in existence by following worldlines of maximal proper time---and the chronometric \emph{potentia} of surrounding matter-energy, understood here not as an additional mathematical quantity in the field equations, but as a conceptual designation for the causal capacity of material sources to modulate the local proper-time interval $ds^2 = g_{\mu\nu}dx^\mu dx^\nu$. Within this relational view, Pseudo-Riemannian geometry emerges a posteriori as the macroscopic mathematical realization required for this chronometric interaction to remain dynamically consistent \cite{Haro2026a}. To develop this program systematically, the remainder of the paper is organized as follows: Sec.~\ref{Zurich} reviews Einstein's struggle in the Z\"urich Notebook and derives the local covariant conservation law directly from fluid kinematics and the Principle of Equivalence without invoking Riemannian geometry; Sec.~\ref{sec:linear_gravity} establishes the linearized wave equations in harmonic coordinates from proper-time free fall and Lorentz boosts; Secs.~\ref{sec:non-linear} and~\ref{sec:I} construct the non-linear completion via the Laplace--Beltrami operator and prove that energy-momentum conservation uniquely fixes the quadratic self-interaction terms into the full Ricci tensor; Sec.~\ref{sec:spin} contrasts this non-perturbative emergence with the spin-2 field-theoretic program; and Sec.~\ref{sec:natural_philosophy} examines the natural-philosophical implications of this chronometric framework, before concluding in Sec.~\ref{sec:conclusions}.

 \

Throughout this work, we adopt  units where $c=1$, and $G$ denotes Newton's gravitational constant.

\section{The Z\"urich notebook and the limits of the Equivalence Principle}\label{Zurich}

In the early stages of his quest for a relativistic theory of gravitation, culminating in the renowned Z\"urich Notebook (1912--1913), Albert Einstein already possessed the fundamental physical ingredients required to build the theory. First, he had incorporated Minkowski's insight that the invariant 
\begin{equation}
ds^2_{\rm Min}:=\eta_{\mu\nu}dx^{\mu}dx^\nu,
\end{equation} 
represented the infinitesimal of proper time \cite{Minkowski1909}. Second, guided by his profound physical intuition regarding the Principle of Equivalence (PE), he had realized that gravity modifies the flow of time \cite{Einstein1911}, and thus alters the structure of the Minkowski invariant $ds^2_{\rm Min}$, which has the general form $ds^2:=g_{\mu\nu}dx^{\mu}dx^\nu$, where $g_{\mu\nu}$ are the \textit{gravitational potentials}.

In addition, he had also realized that the motion of test particles is obtained from the extremization of the proper time $s=\int ds$ \cite{Einstein1912a,Einstein1912b}. From this equation of motion, he had deduced, in a heuristic way, the conservation law for the stress-energy tensor of a dust fluid, 
\begin{equation}
T^{\mu\nu} := \rho u^\mu u^\nu,
\end{equation} 
where $\rho$ is the energy density in the 
comoving frame, which he wrote in the form
\begin{equation}
 \frac{1}{\sqrt{-g}} \partial_\mu \left( \sqrt{-g} T^\mu_\nu \right) - \frac{1}{2} (\partial_\nu g_{\rho\lambda}) T^{\rho\lambda}=0,
\label{Einstein_conservation}
\end{equation}
and, as Marcel Grossmann pointed out in \cite{Grossmann1914}, 
it  can be expressed geometrically in terms of the covariant derivative as $\nabla_\mu T^{\mu}_{\nu} = 0$. 

Finally, the mathematical framework for the wave operator related to the general invariant $ds^2=g_{\mu\nu}dx^\mu dx^\nu$ was available through Eugenio Beltrami's 1868 generalization of the Laplace operator applied to a scalar function $f$, as follows:
\begin{equation}
    \Box_{\rm LB}f:=\frac{1}{\sqrt{-g}}\partial_\mu(\sqrt{-g}g^{\mu\nu}\partial_\nu f).
\end{equation}

By imposing the harmonic coordinate condition, $\partial_\mu (\sqrt{-g} g^{\mu\nu}) = 0$, this operator acquires the simplified form 
\begin{equation}
\Box_{\rm LB} f = g^{\mu\nu}\partial^2_{\mu\nu} f.
\end{equation}

With these tools, the linearized field equations in harmonic coordinates \cite{Straumann2011} naturally emerge  for the perturbation of the Minkowski invariant, namely $h_{\mu\nu}:=g_{\mu\nu}-\eta_{\mu\nu}$,
\begin{equation}\label{lin_eq_reversingtrace}
\Box\left(h_{\mu\nu}-\frac{1}{2}\eta_{\mu\nu}h \right)=-16\pi G T_{\mu\nu}, 
\end{equation}
where $\Box$ denotes the d'Alembert operator and $h$ is the trace of $h_{\mu\nu}$.

When the linearized harmonic condition $\partial_\mu(h^\mu_\nu - \frac{1}{2}\delta^\mu_\nu h) = 0$ is applied, the conservation of the stress-energy tensor in Minkowski spacetime, $\partial_\mu T^\mu_\nu = 0$, is perfectly recovered.

By reversing the trace, this yields the linear wave equation 
\begin{equation}\label{lin_eq}
\Box h_{\mu\nu} = -16\pi G\left(T_{\mu\nu} - \frac{1}{2}\eta_{\mu\nu}T\right),
\end{equation}
where  $T:=\eta^{\mu\nu}T_{\mu\nu}$ is the trace of the stress tensor. 

\subsection{Non-geometric deduction of the conservation law}
\label{subsection:A}
In standard presentations of General Relativity, Einstein's conservation law~(\ref{Einstein_conservation}) is obtained geometrically---either by replacing ordinary derivatives with covariant ones or as a consequence of the contracted Bianchi  identities \cite{Bianchi}. Consistent with our bottom-up perspective, however, we derive this law directly from elementary physical principles, without presupposing a Riemannian geometric structure. In what follows, we show how Eq.~(\ref{Einstein_conservation}) emerges step by step from classical fluid kinematics, special-relativistic energy-momentum equivalence and thermodynamics, and the Principle of Equivalence---first for incoherent dust and subsequently for a general perfect fluid with pressure.

We begin with the non-relativistic continuity equation for a fluid of mass density $\varrho$ and velocity field $\mathbf{v}$,
\begin{equation}
\partial_t\varrho + \nabla\cdot(\varrho\mathbf{v}) = 0,
\end{equation}
where $\nabla := (\partial_x,\partial_y,\partial_z)$ is the nabla operator. This equation follows directly from the Reynolds transport theorem \cite{Gurtin1981}: for any smooth scalar function $f(t,x,y,z)$ and spatial domain $V$,
\begin{equation}
\left.\frac{d}{dt}\int_{\phi_t(V)} f(t,x,y,z)\,dx\,dy\,dz\right|_{t=0} =
\int_V \left(\partial_t f + \nabla\cdot(f\mathbf{v})\right)dx\,dy\,dz,
\end{equation}
where $\phi_t$ denotes the flow generated by the velocity field $\mathbf{v}$. Setting $f = \varrho$ and demanding conservation of mass $\left.\frac{d}{dt}\int_{\phi_t(V)}\varrho\,d^3x\right|_{t=0}=0$ for every arbitrary domain $V$ immediately yields the continuity equation.

Next, we pass to Minkowski spacetime. In a given inertial frame, Lorentz volume contraction relates the coordinate density $\varrho$ to the proper energy density $\rho$ (i.e., the energy density measured in the comoving frame) via $\varrho = \gamma\rho$, where $\gamma := (1 - |\mathbf{v}|^2)^{-1/2}$ (in units $c=1$) is the Lorentz factor. Substituting this relation into the continuity equation gives $\partial_t(\gamma\rho) + \nabla\cdot(\gamma\rho\mathbf{v}) = 0$. Recalling that the four-velocity is $u^\mu = dx^\mu/ds = (\gamma, \gamma\mathbf{v})$, the conservation law takes the manifestly covariant form
\begin{equation}
\partial_\mu(\rho u^\mu) = 0.
\end{equation}

Applying the four-dimensional Reynolds transport theorem along the spacetime flow $\phi_s$ generated by $u^\mu$ over an arbitrary four-dimensional domain $\Omega$ gives
\begin{equation}
\left.\frac{d}{ds}\int_{\phi_s(\Omega)} \rho\,d^4x\right|_{s=0} = \int_\Omega \partial_\mu(\rho u^\mu)\,d^4x = 0,
\end{equation}
where $d^4x \equiv dt\,dx\,dy\,dz$. Note that the integral
\begin{equation}
S_m[\Omega] := \int_{\Omega} \rho\,d^4x
\end{equation}
carries dimensions of action; thus, in Special Relativity, matter conservation can equivalently be expressed as the invariance of this matter integral under the flow generated by the fluid four-velocity, that is,
\begin{equation}
   \left. \frac{d}{ds}S_{ m}[\phi_s(\Omega)]\right|_{s=0}=0.
\end{equation}

Finally, we incorporate gravity following a line of reasoning analogous to Einstein's in the \textit{Entwurf} \cite{Grossmann1914}, but emphasizing an insight that Einstein had already obtained in 1911 from the Principle of Equivalence: gravity affects the rate at which clocks run \cite{Einstein1911}. His celebrated argument involving a uniformly accelerated cabin shows that two light signals emitted at equal time intervals from its ceiling reach the floor at unequal intervals. By the Principle of Equivalence, the same effect must occur in a gravitational field.
Gravity therefore requires a generalization of the Minkowski infinitesimal interval: the flat-spacetime expression \(ds_{\rm Min}\), which describes proper time throughout Minkowski spacetime, must be replaced by an expression that accounts for the influence of the gravitational field. As we shall see, however, the Principle of Equivalence ensures that this generalized interval can always be brought locally to its Minkowskian form in a freely falling frame.

We therefore introduce gravitational potentials through the generalized expression for the infinitesimal interval,
\begin{equation}
ds^2=g_{\mu\nu}dx^\mu dx^\nu.
\end{equation}

This change in the coordinate expression of the proper-time interval has a close conceptual parallel with the Principle of d'Alembert: \textit{The balance between applied and inertial forces vanishes}. As discussed in our previous work \cite{Haro2026b}, the same dynamical balance can take different forms depending on the observer's reference frame. For an observer at rest on the Earth's surface, the equation governing a freely falling particle, with the vertical axis pointing upwards, is
\begin{equation}
-mg-m\ddot z=0.
\end{equation}

For an observer freely falling with the particle, using the coordinates \(\bar z=z+\frac12gt^2\), the same balance becomes
\begin{equation}
-m\ddot{\bar z}=0.
\end{equation}

The gravitational and inertial contributions appearing in the first description are absent from the second, although both observers describe the same physical motion. The underlying dynamical law remains unchanged; what changes is the way its terms are represented in each reference frame.

A closely related distinction arises in the description of proper time. The same invariant interval is expressed through the gravitational potentials \(g_{\mu\nu}\) in general coordinates, while it takes the Minkowskian form in a locally freely falling frame. In both cases, the Principle of Equivalence allows us to distinguish the physical content of a law from its particular coordinate expression. The different mathematical forms do not represent different physical laws, but different descriptions of the same physical situation.

A body in free fall experiences no local gravitational force: its natural motion, understood here in terms of its proper-time evolution, is locally indistinguishable from inertial motion in the absence of gravity. Accordingly, at any given event \(P\), we can introduce  locally comoving, freely falling inertial coordinates
\(X^a=(T,\mathbf X)=(T,X,Y,Z)\), 
where \(T\) measures proper time along the worldline of the freely falling observer, in which the interval takes the Minkowskian form up to terms of second order in the distance from \(P\):
\begin{equation}\label{ds_galilean}
ds^2=[\eta_{ab}+\mathcal O(|X-X_P|^2)]\,dX^a dX^b.
\end{equation}

At the event \(P\), therefore, the local interval reduces to its Minkowskian form, even though its expression in arbitrary coordinates retains the gravitational potentials. This is precisely the distinction we wish to emphasize: gravity modifies the proper-time interval as described in a general frame, while its local influence on free-fall motion can be eliminated by passing to a freely falling frame. The latter does not remove tidal effects, which are reflected in the terms of second order and beyond.

In these locally freely falling coordinates, the local matter integral element is \(\rho\,d^4X\). 
Under a transformation to an arbitrary coordinate system \(x^\mu\), corresponding to a different description of the same physical situation, the measure transforms according to the Jacobian
\begin{equation}
J(P)=\det\left(\frac{\partial X^a}{\partial x^\mu}(P)\right),
\end{equation}
so that
\begin{equation}
\rho\,d^4X=\rho\,|J(P)|\,d^4x.
\end{equation}

Simultaneously, the interval at \(P\) takes the general form
\begin{equation}
ds^2=g_{\mu\nu}(P)\,dx^\mu dx^\nu,
\end{equation}
where the gravitational potentials have the form
\begin{equation}
g_{\mu\nu}(P)=
\eta_{ab}\,
\frac{\partial X^a}{\partial x^\mu}(P)
\frac{\partial X^b}{\partial x^\nu}(P).
\end{equation}

The transformation above makes explicit how the same invariant interval is represented in two different coordinate systems. In the locally freely falling frame, its components reduce at \(P\) to the Minkowski form, whereas in arbitrary coordinates they are encoded in the coefficients \(g_{\mu\nu}\). The gravitational field is therefore reflected in the coordinate representation of the interval, without affecting its invariant character. Tidal effects, which cannot be eliminated throughout a finite region by a change of reference frame, remain physically significant.

It is also important to stress that this construction does not presuppose spacetime curvature or an underlying Riemannian geometric structure. The invariant volume element is obtained by applying the local coordinate transformation and its Jacobian at each event, without requiring a single freely falling coordinate system to cover the entire domain, while the quantities \(g_{\mu\nu}\) arise as the coefficients of the transformed Minkowski interval. They are interpreted here as gravitational potentials, not as components of a pre-existing geometric structure. No affine connection, covariant derivative, or Riemann curvature tensor is introduced as a foundational postulate. The corresponding geometric structures enter only subsequently, as mathematical expressions of relations obtained from physical principles.

Taking determinants on both sides of the metric definition yields $g(P) \equiv \det(g_{\mu\nu}(P)) = -J^2(P)$, whence $|J(P)| = \sqrt{-g(P)}$. Since this relation holds at every event \(P\), the invariant matter integral over any finite four-dimensional domain \(\Omega\) can be expressed in arbitrary coordinates as
\begin{equation}
S_m[\Omega]=\int_\Omega \sqrt{-g}\rho d^4x.
\end{equation}

This construction does not require a single comoving coordinate system covering the entire domain, since the local freely falling frames can be introduced independently at each event, with their descriptions consistently related through the corresponding coordinate transformations.

Requiring the invariance of this matter integral under the four-velocity flow, $\left.\frac{d}{ds}S_m[\phi_s(\Omega)]\right|_{s=0} = 0$, the four-dimensional Reynolds transport theorem immediately yields 
\begin{equation}\label{II.14}
\partial_\mu(\sqrt{-g}\,\rho u^\mu) = 0 \Longrightarrow \frac{1}{\sqrt{-g}}\partial_\mu(\sqrt{-g}\,\rho u^\mu) = 0.
\end{equation}

To complete the derivation, we combine Eq.~(\ref{II.14}) with the free-fall equation of motion parameterized by proper time,
\begin{equation}\label{II.15}
\frac{Du^\nu}{ds} := \frac{du^\nu}{ds} + \Gamma^\nu_{\alpha\beta}u^\alpha u^\beta = 0 \Longrightarrow \frac{Du_\nu}{ds} := \frac{du_\nu}{ds} - \frac{1}{2}(\partial_\nu g_{\alpha\beta})u^\alpha u^\beta = 0,
\end{equation}
where $\Gamma^\nu_{\alpha\beta}$ are the Christoffel symbols obtained from the stationarity of proper time $\delta\int ds = 0$. Multiplying Eq.~(\ref{II.14}) by $u_\nu$ and adding $\rho\,\frac{Du_\nu}{ds} = 0$ gives
\begin{equation}
\frac{1}{\sqrt{-g}}\partial_\mu(\sqrt{-g}\,\rho u^\mu)u_\nu + \rho\frac{Du_\nu}{ds} = 0.
\end{equation}

Identifying the energy-momentum tensor for incoherent dust as $T^\mu_{\nu} = \rho u^\mu u_\nu$, the left-hand side expands exactly into the covariant divergence $\nabla_\mu T^\mu_{\nu}$. 
Here, $\nabla_\mu T^\mu_\nu$ is introduced purely as a compact shorthand for the left-hand side of Eq.~(\ref{Einstein_conservation})---assembled exclusively from the Jacobian factor $\sqrt{-g}$ and the stationarity condition $\delta \int ds = 0$---without presupposing an affine connection or curvature tensor on a manifold.

The extension to a general
adiabatic perfect fluid with pressure $p$, described by \begin{equation}
\label{stress-tensor_perfect_fluid}
T^\mu_{\nu} = (\rho + p)u^\mu u_\nu - p\delta^\mu_\nu,\end{equation} follows the same principles. Because pressure performs mechanical work during fluid expansion or compression, proper energy is no longer conserved along the flow. Instead, applying the four-dimensional Reynolds transport theorem to the conserved particle-number density $n$ gives $\frac{1}{\sqrt{-g}}\partial_\mu(\sqrt{-g}\,n u^\mu) = 0$. Combining this with the first law of thermodynamics \cite{Weinberg1972},
\begin{equation}
d(\rho/n) = -p\,d(1/n) \Longrightarrow d\rho = \frac{\rho+p}{n}dn,
\end{equation}
where we have used that the particle number $N := nV$ within a comoving volume $V$ is constant, yields the energy continuity equation
\begin{equation}
\frac{1}{\sqrt{-g}}\partial_\mu\big[\sqrt{-g}\,(\rho + p) u^\mu\big] - u^\mu\partial_\mu p = 0.
\end{equation}

Furthermore, in Minkowski spacetime, the energy continuity equation derived above from the first law of thermodynamics reads $\partial_\mu(\rho u^\mu) + p\,\partial_\mu u^\mu = 0$. Expanding this equation in an inertial frame $(t,\mathbf{x})$ for an infinitesimal fluid velocity $\vert{}\mathbf{v}\vert{} \ll 1$ (so that $\gamma = 1 + \mathcal{O}(\vert{}\mathbf{v}\vert{}^2)$) and using the identity $p\,\nabla\cdot\mathbf{v} = \nabla\cdot(p\mathbf{v}) - \mathbf{v}\cdot\nabla p$ yields
\begin{equation}
\partial_t\rho + \nabla\cdot\big[(\rho + p)\mathbf{v}\big] = \mathbf{v}\cdot\nabla p.
\end{equation}

This directly identifies $(\rho + p)\mathbf{v}$ as the total energy flux---combining the convective transport of internal energy ($\rho\mathbf{v}$) with the mechanical work done by the pressure ($p\mathbf{v}$)---while $\mathbf{v}\cdot\nabla p$ accounts for the conversion of internal energy into bulk kinetic energy by the pressure gradient. At this point, a fundamental relativistic principle intervenes: the equivalence of mass and energy dictates that any flow of energy inherently constitutes a flow of mass, and therefore a momentum density. By this special-relativistic equivalence between energy flux and momentum density, first established by Max von Laue \cite{Laue1911}, the momentum density to first order in $\mathbf{v}$ must be $\mathbf{g} = (\rho + p)\mathbf{v}$. This reveals a profound physical insight: the inertial mass density of a pressurized fluid is not simply its energy density $\rho$, but its enthalpy density $\rho + p$.

To obtain the corresponding equation of motion, choose a fixed inertial frame $\Sigma_0$ with coordinates $X^a = (T,\mathbf{X})$ in which a given infinitesimal fluid element of volume $\delta V$ is \emph{instantaneously} at rest at $T=0$ (hence $\mathbf{v}(0)=\mathbf{0}$). At neighboring instants $T$, the accelerating element acquires an infinitesimal velocity $\mathbf{v}(T)$ ($|\mathbf{v}|\ll 1$), meaning its momentum is $\delta\mathbf{P} = \mathbf{g}\,\delta V = (\rho+p)\mathbf{v}\,\delta V + \mathcal{O}(|\mathbf{v}|^3)$, while the net pressure force exerted on it is $-(\nabla_{\mathbf{X}} p)\,\delta V$ (arising from the pressure difference across opposite faces of the volume element).
Applying Newton's second law—or, more precisely in its mathematical formulation for continuous media, Euler's equation of motion for fluids
\footnote{Isaac Newton formulated his second law in the \textit{Principia} in verbal terms: ``The change of motion is proportional to the motive force impressed, and takes place along the straight line in which that force is impressed.'' The familiar mathematical expression $\mathbf{F}=m\mathbf{a}$, however, does not appear in the \textit{Principia}. Its explicit analytical formulation is due to Leonhard Euler, who made it a fundamental principle of mechanics in his paper \textit{Découverte d’un nouveau principe de mécanique} (1752). This distinction is not intended to suggest a different physical content, but rather to acknowledge the historical development from Newton's verbal formulation to Euler's mathematical expression, which provided the framework for the systematic formulation of the equations of motion of continuous media.}\cite{Euler}---requires $\frac{d(\delta\mathbf{P})}{dT} = -(\nabla_{\mathbf{X}} p)\,\delta V$. Expanding the time derivative of momentum via the product rule and evaluating at $T=0$ cancels the term proportional to $\mathbf{v}$ (since $\mathbf{v}(0)=\mathbf{0}$), leaving $(\rho+p)\frac{d\mathbf{v}}{dT} = -\nabla_{\mathbf{X}} p$. Since at $T=0$ in $\Sigma_0$ one has $u_a = (1,\mathbf{0})$, $ds = dT$, and $\frac{du_a}{ds} = (0, -\frac{d\mathbf{v}}{dT})$, this relation coincides component by component with the four-vector identity $(\rho + p)\frac{du_a}{ds} = \partial_a p - \frac{dp}{ds}u_a$.

Applying a Lorentz transformation $X^a = \Lambda^a_{\ \nu} x^\nu$ (with $\Lambda^a_{\ \nu} = \partial X^a/\partial x^\nu$) to an arbitrary inertial frame $x^\nu$ then yields the special-relativistic Euler equation,
\begin{equation}\label{SR_Euler}
(\rho + p)\frac{du_\nu}{ds} = \partial_\nu p - \frac{dp}{ds}u_\nu.
\end{equation}

Finally, we extend Eq.~(\ref{SR_Euler}) to an arbitrary coordinate system $x^\mu$ without invoking any ``minimal substitution'' prescription. 
At each event $P$, introduce once again locally comoving, freely falling inertial coordinates $X^a$ in which the infinitesimal proper time has the form~(\ref{ds_galilean}) and the special-relativistic Euler equation holds locally, $(\rho + p)\frac{du_a}{ds}(P) = \partial_a p(P) - \frac{dp}{ds}(P)u_a(P)$.
 Transforming the covelocity $u_a = \frac{\partial x^\nu}{\partial X^a}u_\nu$ along the fluid worldline via the chain rule yields
\begin{equation}
\frac{du_a}{ds}(P) = \frac{\partial x^\nu}{\partial X^a}(P)\frac{du_\nu}{ds}(P) + \frac{\partial^2 x^\nu}{\partial X^a \partial X^b}(P)\frac{dX^b}{ds}(P)\,u_\nu(P).
\end{equation}

Differentiating the potential relation $g_{\mu\nu} = \eta_{ab}\frac{\partial X^a}{\partial x^\mu}\frac{\partial X^b}{\partial x^\nu}$ at $P$ (where $\partial_c g_{ab}(P) = 0$) expresses the second derivatives of the coordinate map directly in terms of the Christoffel symbols already defined in Eq.~(\ref{II.15}), namely \begin{equation}\frac{\partial^2 x^\nu}{\partial X^a \partial X^b}\frac{\partial X^b}{\partial x^\beta}\frac{\partial X^a}{\partial x^\alpha}(P) = -\Gamma^\nu_{\alpha\beta}(P).\end{equation}

Substituting $\frac{dX^b}{ds} = \frac{\partial X^b}{\partial x^\beta}u^\beta$ and relabelling dummy indices therefore gives
\begin{equation}
\frac{du_a}{ds}(P) = \frac{\partial x^\nu}{\partial X^a}(P)\left[\frac{du_\nu}{ds}(P) - \Gamma^\alpha_{\nu\beta}(P)u^\beta(P)u_\alpha(P)\right] = \frac{\partial x^\nu}{\partial X^a}(P)\frac{Du_\nu}{ds}(P),
\end{equation}
whereas the pressure gradient transforms homogeneously as $\partial_a p - \frac{dp}{ds}u_a = \frac{\partial x^\nu}{\partial X^a}(\partial_\nu p - \frac{dp}{ds}u_\nu)$, because the pressure  is a scalar.

Contracting both sides with the inverse Jacobian $\frac{\partial X^a}{\partial x^\mu}(P)$ and relabelling $\mu \to \nu$ yields the relativistic Euler equation,
\begin{equation}\label{Euler_eq_rel}
(\rho + p)\frac{Du_\nu}{ds} = \partial_\nu p - \frac{dp}{ds}u_\nu.
\end{equation}

The preceding derivation also clarifies why the energy-momentum tensor of a perfect fluid takes the form introduced in (\ref{stress-tensor_perfect_fluid}). Its components encode precisely the physical quantities obtained above from independent principles: the proper energy density \(\rho\), the energy flux and momentum density \((\rho+p)\mathbf v\) to first order in the fluid velocity, and the isotropic pressure \(p\). The tensor is therefore not introduced as an independent assumption, but emerges naturally as the mathematical structure that brings together the two physical laws derived separately: energy conservation and the equation of motion. These two laws are then unified in the single condition that the divergence of the energy-momentum tensor vanishes. In fact, multiplying the energy continuity equation by \(u_\nu\) and adding the relativistic Euler equation yields precisely Einstein's conservation law~(\ref{Einstein_conservation}) for a perfect fluid.

\begin{remark}[Variational formulation of the Euler equation and generalized \emph{conatus}]
\label{rem:variational_euler}

For a homentropic perfect fluid (constant specific entropy throughout the fluid), the first law of thermodynamics $d(\rho/n) = -p\,d(1/n)$ 
implies that the specific enthalpy per particle,
\begin{equation}
w := \frac{\rho + p}{n},
\end{equation}
satisfies $dw = n^{-1}dp$ and hence $\partial_\nu w = n^{-1}\partial_\nu p$. Dividing the relativistic Euler equation (\ref{Euler_eq_rel}) by $n$ then brings it into the canonical form

\begin{equation}
\frac{d}{ds}(wu_\nu)
-\frac{1}{2}w(\partial_\nu g_{\alpha\beta})u^\alpha u^\beta
=\partial_\nu w,
\end{equation}
which coincides with the Euler–Lagrange equations of the generalized stationary-action principle \cite{Lichnerowicz1941}
\begin{equation}
\delta\int w\,ds=0,
\end{equation}
(see \cite{Taub1954,Lichnerowicz1967} for the corresponding four-dimensional variational formulation and the interpretation of \(w\) as the fluid index).

This variational principle admits a natural interpretation in terms of a conformal gravitational field. Indeed, introducing the effective gravitational potentials
\begin{equation}
\widetilde{g}_{\mu\nu}:=
\left(\frac{w}{m_0}\right)^2g_{\mu\nu},
\end{equation}
where \(m_0\) denotes a reference particle rest mass,  and noting that ($w>0$) for ordinary
matter,
 the corresponding weighted interval becomes
\begin{equation}
d\widetilde{s}:=\frac{w}{m_0}\,ds,
\end{equation}
which may be interpreted as an enthalpy-weighted proper-time interval, reflecting the modulation of inertial motion by the local thermodynamic state of the medium. The variational principle therefore takes the equivalent form
\begin{equation}
\delta\int d\widetilde{s}=0.
\end{equation}

The enthalpy per particle thus acts as a conformal gravitational field modulating the underlying gravitational potentials \(g_{\mu\nu}\), incorporating the local thermodynamic state of the fluid into its inertial dynamics.

Physically, this provides a unified interpretation of fluid motion in terms of a generalized \emph{conatus}. In the pressureless limit, the first law implies \(d(\rho/n)=0\), so that \(w=\rho/n\) remains constant and the conformal gravitational field reduces to a constant rescaling of the underlying gravitational potentials \(g_{\mu\nu}\). The variational principle thus recovers the inertial perseverance of free matter, irrespective of the magnitude of its internal energy. When pressure is present, however, the enthalpy per particle may vary with the thermodynamic state of the fluid, giving rise to a nontrivial conformal gravitational field that modulates the underlying gravitational potentials. The motion of fluid elements is then governed by the stationarity of the effective interval \(\int d\widetilde{s}\), which may be understood as an enthalpy-weighted proper-time interval: the local thermodynamic state determines the relative dynamical weight of each segment of the worldline, thereby extending the inertial \emph{conatus} of free matter to a fluid subject to internal interactions. In this sense, \(w\) plays the role of an effective refractive index, in direct analogy with Fermat's principle in geometrical optics \cite{Fermat1679}.

\end{remark}

\subsection{Heuristic Tension in the Newtonian Limit}
Despite this initial triumph, Einstein famously abandoned the harmonic gauge and the linearized Ricci tensor. The root of this rejection lay in his interpretation of the Principle of Equivalence, which was the cornerstone of his framework. In his 1911 paper, for instance, Einstein used the Doppler effect in a uniformly accelerated reference frame to infer that clocks at different gravitational potentials run at different rates in a static gravitational field \cite{Einstein1911}. This is the physical insight illustrated by the accelerated-cabin argument discussed above: light signals emitted at equal time intervals from the ceiling reach the floor at unequal intervals. The PE thus allowed Einstein to relate the effects of a gravitational field to those of accelerated motion.

However, the PE is fundamentally a heuristic method for local extrapolation, rather than a global identity. In 1912, Einstein arguably took this identification a step too far: after identifying the invariant associated with a uniformly accelerating frame (the Rindler invariant), he interpreted it as physically equivalent to a frame at rest in a uniform gravitational field \cite{Einstein1912b}. This led him to assume that the spatial sector of the invariant had to remain strictly Euclidean.

This assumption forced the weak-field invariant for a static gravitational field to take the form 
\begin{equation}\label{newtonian_invariant}
ds^2 = (1+2\Phi)dt^2 - dx^2 - dy^2 - dz^2,
\end{equation}
where $\Phi$ is the Newtonian potential, which explicitly violates the linearized harmonic condition. 

Since this specific form of the invariant was non-negotiable for Einstein at the time, he abandoned the linearized wave equation. 

In 1913, he started his collaboration with Marcel Grossmann, who introduced Einstein to the absolute differential calculus developed a few years earlier by Gregorio Ricci-Curbastro and Tullio Levi-Civita \cite{RicciLeviCivita1900}. 

They attempted to find a tensorial generalization of the Poisson equation $\Delta \Phi=4\pi G\rho$, using the Ricci tensor $R_{\mu\nu}$ as a natural candidate. Their starting point was the equation:
\begin{equation}
    R_{\mu\nu}=\kappa T_{\mu\nu},
\end{equation}
where $\kappa$ is a universal constant related to Newton's gravitational constant $G$.

Although we now know that this equation is inherently incompatible with the covariant conservation of the stress-energy tensor $T_{\mu\nu}$, the immediate problem for Einstein was that, upon substituting the invariant (\ref{newtonian_invariant}), the Ricci tensor failed to yield consistently the Poisson equation.
 
Therefore, he concluded that the Ricci tensor could not be the correct candidate for the field equations \cite{Grossmann1914}. Even long before,  he had realized that gravity must also curve the spatial sector---a leap he heuristically justified by combining the PE with the Ehrenfest paradox of a rotating disk \cite{Ehrenfest1909}---he 
remained 
tethered to this overly restrictive weak-field ansatz for at least two more years.

\subsection{The 1915 Resolution and the Pseudo-Tensor}

It was only during the intense weeks of November 1915 that Einstein returned to the Ricci tensor. His successful computation of the anomalous perihelion advance of Mercury provided the decisive breakthrough, reassuring him that his mathematical trajectory was sound. To tame the daunting algebraic complexity of the Ricci tensor, he temporarily restricted his analysis to unimodular coordinate systems ($\sqrt{-g} = 1$). On November 25, 1915, he submitted the field equations for a material fluid in the form \cite{Einstein1915}:
\begin{equation}
R_{\mu\nu} = \kappa \left(T_{\mu\nu} - \frac{1}{2}g_{\mu\nu}T\right) \qquad\mbox{with}\qquad
R_{\mu\nu}=\partial_{\alpha}\Gamma_{\mu\nu}^\alpha -\Gamma_{\mu\beta}^\alpha \Gamma_{\nu\alpha}^\beta,
\end{equation}
where $T$ represents the trace of the energy-momentum tensor and  $\kappa\equiv 8\pi G$.

Crucially, 
although the Bianchi identities were already known \cite{Bianchi}, Einstein had not yet incorporated their contracted form into his reasoning as the geometrical mechanism ensuring covariant energy-momentum conservation.
For this reason,
he struggled to harmonize his geometric equations with matter dynamics. Guided by the paradigm of classical electrodynamics, he reasoned that the theory must possess an ordinary conservation law combining both matter and field contributions. By enforcing the local conservation requirement $\nabla_\mu T^{\mu}_{\nu} = 0$ alongside his proposed field equations, he constructed a standard continuity equation:
\begin{equation}
\partial_\mu(T^\mu_\nu + t^\mu_\nu) = 0.
\end{equation}

This formulation necessitated the introduction of $t^\mu_\nu$, a non-tensorial quantity (pseudo-tensor) that he interpreted as the intrinsic energy-momentum of the gravitational field itself.

In his 1916 comprehensive treatise, \textit{The Foundation of the General Theory of Relativity} \cite{Einstein1916}, Einstein formalized this derivation. Remaining within the unimodular gauge, he initially examined vacuum space, where the vanishing of the Ricci tensor led after algebraic manipulation to:
\begin{equation}
 \partial_{\mu} t^{\mu}_{\nu}=0,
\end{equation}
with the gravitational energy-momentum pseudo-tensor $t^{\mu}_{\nu}$ defined explicitly as:
\begin{eqnarray}
\kappa t^{\mu}_{\nu}
\equiv
\frac{1}{2}\delta_{\nu}^{\mu} g^{\alpha\sigma}
\Gamma^{\lambda}_{\alpha\beta}\Gamma^{\beta}_{\sigma\lambda}
-
g^{\alpha\sigma}\Gamma^{\mu}_{\alpha\beta}\Gamma^{\beta}_{\sigma\nu}.
\end{eqnarray}

Drawing a direct analogy to Maxwellian field theory, Einstein identified this object as the stress-energy density of the gravitational field.

Using this definition, the vacuum field equations could be rewritten as:
\begin{eqnarray}
\partial_{\alpha}(g^{\nu\beta}\Gamma_{\mu\beta}^{\alpha})
=
\kappa\left(
t^{\nu}_{\mu}
-
\frac{1}{2} \delta^{\nu}_{\mu} t
\right),
\qquad \mbox{with} \quad g=-1,
\end{eqnarray}
where $t\equiv t^{\alpha}_{\alpha}$ denotes the trace of the gravitational pseudo-tensor.

This structural form provided the essential stepping stone for coupling the field to matter. Einstein's physical reasoning was as elegant as it was compelling: the gravitational field cannot be sourced by matter energy alone, nor purely by field energy, but rather by the total energy-momentum of the interacting system. This insight led directly to the complete field equations:
\begin{eqnarray}\label{Einstein_field}
\partial_{\alpha}(g^{\nu\beta}\Gamma_{\mu\beta}^{\alpha})
=
\kappa\left(
(t^{\nu}_{\mu}+T^{\nu}_{\mu})
-
\frac{1}{2} \delta^{\nu}_{\mu}(t+T)
\right),
\qquad \mbox{with} \quad g=-1,
\end{eqnarray}
which intrinsically embodies the total energy-momentum conservation law:
\begin{equation}
\partial_{\mu}(t^{\mu}_{\nu}+T^{\mu}_{\nu})=0,
\end{equation}
dynamically matching the local covariant conservation of matter, $\nabla_{\mu}T^{\mu}_{\nu}=0$.

Finally, by shifting the pseudo-tensor terms back to the left-hand side, Equation~(\ref{Einstein_field}) yields its final, fully general expression:
\begin{eqnarray}
\partial_{\alpha}\Gamma_{\mu\nu}^{\alpha}
-
\Gamma^{\alpha}_{\mu\beta}\Gamma^{\beta}_{\nu\alpha}
=
\kappa\left(
T_{\mu\nu}
-
\frac{1}{2} g_{\mu\nu} T
\right) \quad \mbox{with} \quad g=-1
\quad \Longleftrightarrow \quad 
R_{\mu\nu}
=
\kappa\left(
{T}_{\mu\nu}
-
\frac{1}{2}{g}_{\mu\nu} T
\right),
\end{eqnarray}
recovering the fully covariant Einstein field equations.

A final remark is in order: this historical trajectory highlights a key epistemological distinction. Einstein did not strictly derive General Relativity from an underlying wave-theoretic framework; instead, constrained by unimodular coordinates, he postulated candidate geometric operators and adjusted them to satisfy energy conservation via gravitational pseudo-tensors. In contrast, our approach demonstrates that spacetime geometry need not be reified \textit{a priori}. By grounding the construction strictly in a non-linear wave equation within harmonic coordinates and enforcing local matter conservation, the metric geometry emerges naturally as the macroscopic realization of gravitational dynamics.

\section{Linear gravity}
\label{sec:linear_gravity}
Having demonstrated in the preceding sections that gravitation operates by distorting the flow of proper time, we can now formulate the core of our program. Our objective is to deduce how material sources modify the proper-time landscape for a test particle, and how the particle dynamically reacts by following its \textit{conatus}—seeking the worldline that locally maximizes its proper time, governed by the variational stationarity condition $\delta\int ds=0$. This framework is constructed through the following steps \cite{Haro2026,Haro2026a}:

\begin{enumerate}
\item \textbf{Static Matter:} We reframe the universal law of Galilean free fall
\cite{Galileu} using the intrinsic proper time of the falling body. By
demanding that this motion be strictly independent of the particle's
velocity, we derive the alteration of proper time due to stationary masses.

    \item \textbf{Moving Matter:} We apply kinematic Lorentz transformations to the resulting static invariant\footnote{This procedure has a remarkable early precedent in Johannes Droste's 1916 doctoral thesis \cite{Droste1916}. In Chapter III, § 1, particularly in § 51, Droste considers a spherically symmetric gravitational center moving at constant velocity relative to the observer. Starting from the static line element obtained in isotropic coordinates, he applies a Lorentz transformation (Eq. 67) to derive the corresponding expression in the moving frame, subsequently expanding the result to obtain the velocity-dependent contributions to the gravitational field (Eqs. 68–70). This provides a clear historical precedent for the central idea adopted here: constructing the gravitational field of moving sources by Lorentz-transforming the corresponding static field. The present treatment develops this strategy within a linearized framework and makes explicit the consistency of superposition at first order.}. This yields, within the weak-field approximation, the proper-time structure generated by masses in motion.
\end{enumerate}

Focusing on the first stage, we express the universality of Galilean free fall exclusively through the proper time ticked by a clock attached to the massive particle. Specifically, we introduce a core dynamical postulate: in a uniform gravitational field, the particle's acceleration parameterized by its proper time is exactly
\begin{equation}\label{eq-freefall}
\frac{d^2z}{ds^2}=-g,
\end{equation}
where here $g$ denotes the magnitude of the constant uniform gravitational acceleration (not to be confused with the determinant of the gravitational potentials $g = \det(g_{\mu\nu})$).

We emphasize that this is not a trivial substitution of the temporal variable in the Newtonian formula $d^2z/dt^2=-g$. Instead, it represents a profound physical shift, linking the inertial motion directly to the object's internal clock.

This conceptual shift dictates that the time parameter governing inertia must not belong to an external reference frame. A particle inherently tracks its own proper time, rendering its intrinsic temporal evolution independent of the observer. While Newtonian physics assumes a universal absolute time, relativity dictates that coordinate time is observer-dependent. Observers situated at different gravitational potentials (such as at the top and bottom of a tower) will record different elapsed times for the exact same falling object due to gravitational time dilation. 

Consequently, stationary observers at different locations provide varying temporal descriptions of the descent. However, the particle measures its own fall using a unique intrinsic parameter: its proper time $s$. Equation (\ref{eq-freefall}) models the vertical coordinate $z$---a spatial measure shared among stationary observers---evolving strictly with respect to this internal clock. Using proper time is therefore a physical necessity, anchoring the temporal measure of inertia to the physical system itself rather than to the coordinate choice of a distant observer.

Next, under initial conditions $\frac{dx}{ds}=\frac{dy}{ds}=0$ which ensures that the test particle moves in the $z$-direction, 
we determine the specific invariant line element that, when subjected to the extremization principle $\delta\int ds=0$, yields equation (\ref{eq-freefall}) exactly.

Assuming spatial isotropy, we propose the following ansatz for the invariant:
\begin{equation}
    ds^2=A(z)dt^2-B(z)d{\bf r}\cdot d{\bf r},
\end{equation}
where ${\bf r}=(x,y,z)$. Applying the stationarity condition $\delta \int ds=0$, the equation of motion parameterized by proper time becomes:
\begin{equation}
    \frac{d^2 z}{ds^2}=-\frac{A'+B'\left(\frac{dz}{dt}\right)^2}{2B\left(A-B\left(\frac{dz}{dt}\right)^2\right)}
    =-\frac{A'+(AB)'\left(\frac{dz}{ds}\right)^2}{2AB},
\end{equation}
with the prime denoting differentiation with respect to the coordinate $z$.

Crucially, enforcing the condition that the equation of motion must be entirely independent of the particle's velocity requires $(AB)'=0$. This implies $B=a/A$ for some constant $a$. Because the gravitational potential vanishes at $z=0$, the invariant must smoothly reduce to the flat Minkowski invariant, thereby fixing $a=1$.

This simplifies the dynamics to:
\begin{equation}
     \frac{d^2 z}{ds^2}=-A'/2.
\end{equation}

To recover the free-fall postulate (\ref{eq-freefall}), we require $A(z)=k+2gz$. Applying the flat-space boundary condition at $z=0$ once more demands $k=1$, yielding:
\begin{equation}
    A(z)=1+2gz.
\end{equation}

Unlike the kinematics of a uniformly accelerated frame in empty space (Rindler coordinates), where the spatial sections remain flat and the effective acceleration exhibits a mild velocity dependence, the strict enforcement of Galilean free fall independent of velocity in a stationary coordinate system dynamically enforces a non-trivial scaling of spatial distances. Specifically, the condition $(AB)' = 0$ dictates that the spatial components must scale precisely as the inverse of the temporal lapse ($A(z)$). This directly leads to the structural form of the conformastatic invariant:
\begin{equation}\label{eq-metric-static}
ds^2 = \left(1+{2gz}\right)dt^2 - \left(1+{2gz}\right)^{-1} d{\bf r}\cdot d{\bf r}, 
\end{equation}
which in the weak-field approximation $|gz|\ll 1$, becomes
\begin{equation}\label{eq-metric-static1}
ds^2 = \left(1+{2gz}\right)dt^2 - \left(1-{2gz}\right) d{\bf r}\cdot d{\bf r}. 
\end{equation}

It is imperative to highlight that no underlying spacetime geometry has been postulated during this process. The invariant  emerges strictly from a dynamical kinematic law (free fall) combined with the proper-time variational principle within a spatially isotropic framework. The interpretation of this invariant as a curved geometry arises only \textit{a posteriori}.

\subsection{Moving sources}

Our starting point is the infinitesimal proper-time interval derived for a strictly uniform gravitational field by enforcing free-fall kinematics (\ref{eq-metric-static}). We now evaluate how moving, pressureless matter distributions (dust) alter the background Minkowski invariant within the weak-field regime.

In this approximation, we systematically restrict our analysis to terms linear in the gravitational perturbation. This truncation has both a physical and a mathematical justification: to extend a static line element to moving sources, we apply Lorentz transformations to the elementary static perturbations. While a Lorentz boost remains mathematically well-defined as a coordinate transformation on the Minkowski background, the linear superposition of independently boosted elementary contributions is valid only at first order. At higher orders, non-linear combinations of the perturbations encode the self-interaction of the gravitational field, rendering simple linear superposition invalid.

Transitioning from the weak-field limit of a strictly uniform acceleration field to an arbitrary static Newtonian potential \(\Phi(\mathbf{r})\) still requires a phenomenological step, as the spatial variation of a general static field cannot be deduced purely from a uniform acceleration profile. Following Einstein's 1912 heuristic of extending a uniform field to a general static one \cite{Einstein1912a, Einstein1912b}, we therefore adopt, as a provisional working hypothesis, the following generalization of the linearized static line element associated with a uniform gravitational field, expressed in the rest frame \(\Sigma\) of the sources (\ref{eq-metric-static1}):
\begin{equation}\label{conformastat_weak}
ds^2 = \bigl(1+2\Phi(\mathbf{r})\bigr)\,dt^2
-\bigl(1-2\Phi(\mathbf{r})\bigr)\,d\mathbf{r}\cdot d\mathbf{r},
\end{equation}
where \(|\Phi(\mathbf{r})|\ll 1\) and the potential \(\Phi\) satisfies the classical Poisson equation \(\Delta\Phi=4\pi G\rho\), with \(\rho\) denoting the rest-frame mass-energy density, in units where \(c=1\).

Consider a new frame $\Sigma'$ moving with a constant three-velocity $-{\bf v}'$ with respect to $\Sigma$. Equivalently, the rest frame $\Sigma$ moves with velocity 
\begin{eqnarray}
   {\bf v}'=\frac{d{\bf r}'}{dt'}
\end{eqnarray}
relative to $\Sigma'$. Using the Lorentz factor
\begin{eqnarray}
   \gamma=\frac{1}{\sqrt{1-|{\bf v}'|^2}} ,
\end{eqnarray}
the standard coordinate transformation that preserves the Minkowski interval is given  by \cite{Jackson1999}:
\begin{eqnarray}
  {\bf r}=\gamma\left[
    \frac{1}{\gamma}{\bf r}'+\frac{\gamma}{1+\gamma}({\bf v}'\cdot {\bf r}'){\bf v}'
    -{\bf v}'t'
    \right],\qquad 
    t=\gamma(t'-{\bf v}'\cdot {\bf r}').
\end{eqnarray}

Applying this transformation, the invariant in $\Sigma'$ generated by the moving source takes the form:
\begin{eqnarray}\label{LORENTZ}
ds^2 &=&
(1 - 2{\Phi}' + 4\Upsilon')\, dt'^2
- 8\, {\mathbf{N}'} \cdot d\mathbf{r}'\, dt'
- (1 - 2{\Phi}')\, d\mathbf{r}' \cdot d\mathbf{r}'
+ 4\, d\mathbf{r}' \cdot \mathbf{U}' \cdot d\mathbf{r}',
\end{eqnarray}
where the transformed potential ${\Phi}'(t',{\bf r}')\equiv\Phi({\bf r}(t',{\bf r}'))$ now identically satisfies Abraham's wave equation \index{Abraham's equation}\cite{Abraham,Nordstrom}:
\begin{eqnarray}
\Box {\Phi}' = -4\pi G {\rho}' \quad \mbox{with}\quad 
{\rho}'(t',{\bf r}')=\rho({\bf r}(t',{\bf r}')).
\end{eqnarray}

The Lorentz-transformed static solution naturally satisfies this wave equation. For an eternally uniformly moving source---as is the case for dust in Minkowski spacetime---this solution coincides with both the retarded and advanced potentials; the distinction between these boundary conditions becomes physically relevant only for genuinely time-dependent source histories.

Consequently, for an eternally uniformly moving source, we can identify the boosted potential $\Phi'(t',{\bf r}')$ with its retarded representation:
\begin{eqnarray}
{\Phi}^{\rm ret}(t',{\bf r}') = -G\int \frac{{\rho}'(t'_{\rm r},\bar{\bf r})}{|{\bf r}'-\bar{\bf r}|}\, d\bar{V},
\end{eqnarray}
where $t'_{\rm r} = t' - |{\bf r}'-\bar{\bf r}|$ is the retarded time. Because the source velocity ${\bf v}'$ is constant, it factors outside the integral, naturally coupling to the scalar, vector, and symmetric tensor potentials via:
\begin{eqnarray}
    \Upsilon' = \frac{{\Phi}^{\rm ret}}{1-|{\bf v}'|^2}, \qquad
    {\bf N}' = \Upsilon' {\bf v}', \qquad
    \mathbf{U}' = \Upsilon' ({\bf v}' \otimes {\bf v}'),
\end{eqnarray}
where
 $\otimes$ denotes the Gibbs dyadic product. For a single velocity, the associated quadratic form trivially reduces to $d\mathbf{r}' \cdot \mathbf{U}' \cdot d\mathbf{r}' = \Upsilon'({\bf v}' \cdot d\mathbf{r}')^2$; however, retaining the full tensorial structure is essential for the macroscopic superposition of arbitrary velocity fields discussed below.

These potentials independently satisfy their respective sourced wave equations:
\begin{eqnarray}
\Box \Upsilon' = -\frac{4\pi G {\rho}'}{1-|{\bf v}'|^2}, \qquad
\Box {\bf N}' = -\frac{4\pi G {\rho}'}{1-|{\bf v}'|^2} {\bf v}', \qquad
\Box \mathbf{U}' = -\frac{4\pi G {\rho}'}{1-|{\bf v}'|^2} ({\bf v}' \otimes {\bf v}').
\end{eqnarray}

Dropping the primes for notational simplicity, the invariant in the generic observer's frame reads:
\begin{eqnarray}\label{LORENTZ1}
ds^2 &=&
(1 - 2{\Phi}^{\rm ret} + 4\Upsilon)\, dt^2
- 8\, {\mathbf{N}} \cdot d\mathbf{r}\, dt
- (1 - 2{\Phi}^{\rm ret})\, d\mathbf{r} \cdot d\mathbf{r}
+ 4\, d\mathbf{r} \cdot \mathbf{U} \cdot d\mathbf{r}.
\end{eqnarray}

We now extend this construction to $N$ material objects with distinct masses and constant velocities. We note that the invariant (\ref{LORENTZ1}) naturally splits into a flat background and a perturbation, $ds^2=ds^2_{\rm Min}+ds^2_{\rm perturbed}$, where:
\begin{eqnarray}\label{perturbed0}
  ds^2_{\rm perturbed}=
(- 2{\Phi}^{\rm ret} + 4\Upsilon)\, dt^2
- 8\, {\mathbf{N}} \cdot d\mathbf{r}\, dt
+ 2{\Phi}^{\rm ret}\, d\mathbf{r} \cdot d\mathbf{r}
+ 4\, d\mathbf{r} \cdot \mathbf{U} \cdot d\mathbf{r}.
\end{eqnarray}

In the comoving frame of the source (denoted by subscript "${\rm c}$"), this perturbation simplifies to the static weak-field form (see Eq.~(\ref{conformastat_weak})):
\begin{eqnarray}\label{perturbed}
ds^2_{\rm perturbed} = 2\Phi({\bf r}_{\rm c})(dt_{\rm c}^2+ d{\bf r}_{\rm c}\cdot d{\bf r}_{\rm c}) \quad \mbox{with}\quad \Delta\Phi=4\pi G\rho.
\end{eqnarray}

For $N$ material objects, the total metric perturbation can therefore be rigorously decomposed into a linear sum of elementary contributions, each evaluated in the inertial coordinates where its respective source is at rest:
\begin{eqnarray}
ds^2_{\rm perturbed}
=
\sum_{k=1}^N
2\Phi_k({\bf r}_k)
\left(
dt_k^2+d{\bf r}_k\cdot d{\bf r}_k
\right),
\end{eqnarray}
where $k=1,\ldots,N$ labels the sources and $\Delta\Phi_k=4\pi G\rho_k$.

Each term represents an independent physical contribution to the infinitesimal proper time. By applying the specific Lorentz boost associated with the constant velocity ${\bf v}_k$ of each source, we express all contributions in the common coordinate system $(t, \mathbf{r})$ of the observer. Since the wave equation is linear, the total macroscopic potentials are strictly the superpositions of the individual boosted fields:
\begin{eqnarray}
    \Phi^{\rm ret}=\sum_{k=1}^N\Phi_k^{\rm ret}, \qquad
    \Upsilon=\sum_{k=1}^N \Upsilon_k, \qquad
    \mathbf{N}=\sum_{k=1}^N \mathbf{N}_k, \qquad
    \mathbf{U}=\sum_{k=1}^N \mathbf{U}_k.
\end{eqnarray}

Taking the continuum limit $N\rightarrow\infty$ naturally yields the macroscopic retarded potentials for a continuous moving matter distribution with a velocity field $\mathbf{v}(t,\mathbf{r})$. Crucially, because the fluid velocity now varies in space and time, it cannot be algebraically factored out of the differential operators. Instead, the kinematics remain intrinsically bound within the retarded source integrals. For instance, the macroscopic spatial tensor potential is rigorously given by:
\begin{eqnarray}
\mathbf{U}(t,{\bf r}) = -G\int \left[ \frac{\rho}{1-|{\bf v}|^2} ({\bf v} \otimes {\bf v}) \right]_{\rm ret} \frac{d\bar{V}}{|{\bf r}-\bar{\bf r}|},
\end{eqnarray}
which identically satisfies the tensorial wave equation:
\begin{eqnarray}
\Box \mathbf{U} = -\frac{4\pi G \rho}{1-|{\bf v}|^2} ({\bf v} \otimes {\bf v}).
\end{eqnarray}

By expressing the coefficients of the resulting invariant as a general perturbation of the Minkowski background, $g_{\mu\nu}=\eta_{\mu\nu}+h_{\mu\nu}$, one can straightforwardly verify (see \cite{Haro2026,Haro2026a} for details) that the perturbation $h_{\mu\nu}$ derived here identically satisfies the standard linear wave equation (\ref{lin_eq}).

\

Finally, a crucial conceptual feature of this bottom-up derivation deserves emphasis: the harmonic coordinate condition is not introduced here as an arbitrary mathematical gauge choice, but emerges naturally as a dynamical consistency requirement, once the appropriate physical boundary conditions are imposed. Defining the trace-reversed perturbation $\bar{h}_{\mu\nu} = h_{\mu\nu} - \frac{1}{2}\eta_{\mu\nu}h$, the linearized wave equation (\ref{lin_eq_reversingtrace}) takes the compact form
\begin{equation}
\Box \bar{h}_{\mu\nu} = -16\pi G T_{\mu\nu}.
\end{equation}

Taking the flat-space divergence of this equation and raising indices, we obtain
\begin{equation}
\Box\left( \partial_\mu \bar{h}^{\mu\nu} \right) = -16\pi G \partial_\mu T^{\mu\nu}.
\end{equation}

Enforcing the physical conservation of the matter stress-energy tensor in Minkowski spacetime, $\partial_\mu T^{\mu\nu} = 0$, immediately yields
\begin{equation}
\Box \left( \partial_\mu \bar{h}^{\mu\nu} \right) = 0.
\end{equation}

The divergence of the trace-reversed perturbation therefore satisfies the source-free wave equation. To select the physically relevant solution, we impose asymptotic flatness together with retarded boundary conditions, excluding any independent source-free contribution to this divergence. Equivalently, its homogeneous initial data are taken to vanish in the remote past. Under this prescription, the homogeneous solution is uniquely fixed to zero, yielding the linear harmonic condition
\begin{equation}
\partial_\mu \bar{h}^{\mu\nu} = 0.
\end{equation}

This restriction applies only to the divergence of the perturbation and does not exclude outgoing gravitational radiation generated by the matter source itself.

In this gauge, the explicit structure of (\ref{LORENTZ1}) makes the connection between linearized and Newtonian gravity completely transparent: it presents the gravitational field as the relativistic, retarded extension of the Newtonian potential via velocity-coupled scalar, vector, and tensor potentials that obey Lorentz-type continuity relations and smoothly recover the static weak-field Newtonian invariant in the low-velocity limit.

Consequently, when extending the wave equation to the full non-linear regime in Sec.~\ref{sec:non-linear}, adopting its exact non-linear completion $\partial_\mu(\sqrt{-g}\,g^{\mu\nu}) = 0$ (i.e., the full harmonic gauge) is not an \emph{ad hoc} restriction, but the direct continuation of the coordinate condition mandated by energy-momentum conservation at linear order.

\

\noindent\textbf{A final remark is in order.} Our linearized construction can be extended from pressureless dust to a general perfect fluid supporting pressure $p$ without presupposing the form of the stress-energy tensor. To do so, one must introduce two distinct effective densities: the active gravitational energy density $\rho + 3p$---which emerges directly from combining Newtonian gravitation with the first law of thermodynamics \cite{Haro2026}---and the effective inertial energy density (the enthalpy density) $\rho + p$, which governs momentum transport via the thermodynamic energy flux and relativistic momentum-energy equivalence. Furthermore, an essential kinematic distinction arises at the level of the fluid trajectories in Minkowski spacetime. For pressureless dust, flat-space conservation $\partial_\mu T^\mu_{\nu} = 0$ implies $\frac{du_\nu}{ds} = 0$, so individual fluid elements follow straight inertial worldlines; their velocities are therefore constant, and the velocity at the retarded event coincides with the present velocity. In contrast, when pressure is present, as shown in \ref{subsection:A}, the flat-space conservation law yields the special-relativistic Euler equation $(\rho + p)\frac{du_\nu}{ds} = \partial_\nu p - \frac{dp}{ds}u_\nu$. Because internal pressure gradients accelerate the fluid elements even in flat spacetime, they no longer follow straight worldlines. Consequently, the velocity entering the local Lorentz transformation from the instantaneous rest frame of each source element must be explicitly evaluated at the retarded event on its past light cone (see \cite{Haro2026a, Haro2026, Haro2026b} for details).

\section{The non-linear wave equation in harmonic coordinates}
\label{sec:non-linear}

Our starting point is the linearized wave equation (\ref{lin_eq}),
in which the Minkowski metric $\eta_{\mu\nu}$ enters the source
through the trace-reversed energy-momentum tensor. To construct
its tensorial non-linear extension, we promote the background
metric $\eta_{\mu\nu}$ to the dynamical metric $g_{\mu\nu}$,
thereby replacing the linearized source by
\begin{equation}
    T_{\mu\nu}-\frac{1}{2}g_{\mu\nu}T.
\end{equation}
This is the natural tensorial generalization of the linearized
source within the metric framework considered here.

Furthermore,  the stress-energy tensor obeys the covariant conservation law
$\nabla_{\mu}T^{\mu}_{\nu}=0$, derived in Subsection
\ref{subsection:A} from matter conservation and the proper-time
dynamics of particle trajectories. The Levi-Civita connection thus
provides a compact geometric representation of a relation whose
physical content was established independently. In the weak-field
limit, this conservation law reproduces the classical continuity and
Euler equations.

Next, we must generalize the left-hand side of (\ref{lin_eq}). In the linear regime, the standard d'Alembertian dictates wave propagation strictly tied to the Minkowski proper time. However, because the presence of matter dynamically alters the physical differential proper time invariant $ds$, the wave operator must inherently adapt to this new chronometric reality. Therefore, we replace the flat d'Alembertian with the Laplace-Beltrami operator. Crucially, in our framework, this replacement is not introduced as an a priori geometric postulate of a curved manifold, but rather as a physically motivated minimal coupling. While not mathematically inevitable from pure algebra without presupposing a metric structure, it emerges as the most natural coordinate-covariant scalar wave operator dynamically synchronized with the altered proper time. 

Furthermore, to this operator we must add a non-linear correction that reproduces the self-interaction of the gravitational field. Physically, since the gravitational field carries energy and must act as its own source, this self-interaction term must mimic the energy density of a continuous field. In classical field theory, energy density is universally constructed from the square of the field's gradients.

In addition, dimensional consistency with the second-order wave operator dictates that this non-linear correction must have the same physical dimensions ($L^{-2}$). 
Although more general non-linear frameworks involving higher-order derivatives---such as $f(R)$ or higher-curvature modified gravity  \cite{Felice,Nojiri}---could be contemplated, they generically alter the differential order of the system, introducing extra propagating degrees of freedom and, in some cases, potential Ostrogradsky instabilities \cite{Ostrogradsky,Woodard}. 
We therefore adopt the minimal second-order completion in which the self-interaction term $U_{\mu\nu}$ is quadratic in the first derivatives of the metric, mirroring the gradient-squared energy density of classical field theory while preserving a well-posed, strictly second-order hyperbolic wave equation.
Then, the full non-linear wave equation in harmonic coordinates takes the form:
\begin{equation}\label{nonlinear_wave_eq}
\Box_{\rm LB}g_{\mu\nu} + U_{\mu\nu} = -16\pi G\left( T_{\mu\nu}-\frac{1}{2}g_{\mu\nu}T \right),
\end{equation}
where $U_{\mu\nu}$ is a pseudo-tensor strictly quadratic in the first derivatives of the gravitational potentials  $g_{\mu\nu}$.

We investigate the structural consistency of the gravitational field equations in full, non-linear General Relativity without invoking the Riemann or Ricci tensors a priori. 

We begin with the trace-reversed field equations where the scalar matter source has been explicitly eliminated:
\begin{equation}
\Box_{\rm LB} g_{\mu\nu} + U_{\mu\nu} - \frac{1}{2}g_{\mu\nu}\big(\operatorname{Tr}(\Box_{\rm LB} \mathfrak{g}) + U\big) = -16\pi G T_{\mu\nu},
\label{fieldeq}
\end{equation}
where $U \equiv \operatorname{Tr}(U) = g^{\alpha\beta}U_{\alpha\beta}$ is the trace of $U_{\mu\nu}$, and $\operatorname{Tr}(\Box_{\rm LB} \mathfrak{g}) \equiv g^{\mu\nu}\Box_{\rm LB}g_{\mu\nu}$.

Imposing compatibility with local energy-momentum conservation, $\nabla_\mu T^{\mu}_{\nu} = 0$, the differential consistency condition requires the vanishing of the divergence of the total field equations. Here, once again, consistent with our operational framework, the symbol $\nabla_\mu$ must not be read as an \textit{a priori} geometric connection on a curved manifold, but rather as the purely dynamical differential operator derived in Subsection~\ref{subsection:A} from free-fall particle trajectories and local energy-momentum balance---in the same physical spirit that originally guided Einstein's \textit{Entwurf} formulation. Although the individual wave-operator piece and the self-interaction block $U^\mu_\nu-\frac{1}{2}\delta^\mu_\nu U$
are separately non-tensorial (pseudo-tensorial), their combination must assemble into a true tensor in order to be consistently identified with the matter source;
 hence, by virtue of the linearity of the operator $\nabla_\mu$ evaluated component-wise in harmonic coordinates, the consistency requirement splits cleanly into:
\begin{equation}
\nabla_\mu \left( g^{\mu\alpha}\Box_{\rm LB} g_{\alpha\nu} - \frac{1}{2}\delta^\mu_\nu g^{\alpha\beta}\Box_{\rm LB} g_{\alpha\beta} \right) + \nabla_\mu \left( U^\mu_\nu - \frac{1}{2}\delta^\mu_\nu U \right) = 0.
\label{master_consistency}
\end{equation}

Let us define the second-order divergence piece as ${\mathcal{I}}_\nu$:
\begin{equation}
{\mathcal{I}}_\nu \equiv \nabla_\mu \left( g^{\mu\alpha}\Box_{\rm LB} g_{\alpha\nu} - \frac{1}{2}\delta^\mu_\nu g^{\alpha\beta}\Box_{\rm LB} g_{\alpha\beta} \right).
\end{equation}

As shown at the end of Sec.~\ref{sec:linear_gravity}, consistency between the linearized wave equation and energy-momentum conservation ($\partial_\mu T^{\mu}_{\nu}=0$) already compels the coordinates to obey the linearized harmonic condition. Extending this dynamical constraint to the full non-linear regime naturally selects the exact harmonic coordinate condition (de Donder--Hilbert gauge) \cite{deDonder1921}:
\begin{equation}
\partial_\mu \left( \sqrt{-g}\, g^{\mu\sigma} \right) = 0,
\label{harmonic}
\end{equation}
which allows us to evaluate ${\mathcal{I}}_\nu$ explicitly, show the identical cancellation of all third-order derivatives ($\partial^3 g$), extract the precise cubic residual vector, namely $Q_\nu$, and solve for $U^\mu_\nu$ via a complete pseudo-tensorial ansatz.

\section{Exact Evaluation of ${\mathcal{I}}_\nu$}
\label{sec:I}

To compute the covariant divergence of $E^\mu_\nu := g^{\mu\alpha}\Box_{\rm LB} g_{\alpha\nu} - \frac{1}{2}\delta^\mu_\nu g^{\alpha\beta}\Box_{\rm LB} g_{\alpha\beta}$, noting that the associated pseudo-tensor $E_{\mu\nu} := \Box_{\rm LB} g_{\mu\nu} - \frac{1}{2}g_{\mu\nu} g^{\alpha\beta}\Box_{\rm LB} g_{\alpha\beta}$ (and hence $E^{\rho\lambda}$) is symmetric, we employ the standard Landau--Lifshitz identity \cite{Landau}:
\begin{equation}
\nabla_\mu E^\mu_\nu = \frac{1}{\sqrt{-g}} \partial_\mu \left( \sqrt{-g} E^\mu_\nu \right) - \frac{1}{2} (\partial_\nu g_{\rho\lambda}) E^{\rho\lambda},
\label{landau_id}
\end{equation}
which coincides with the expression derived heuristically by Einstein (\ref{Einstein_conservation}) without resorting to geometry, as we have already explained.

When expanding this full covariant derivative under the harmonic coordinate condition, the connection term $-\frac{1}{2}(\partial_{\nu}g_{\rho\lambda})E^{\rho\lambda}$ identically cancels the partial derivatives acting on the square root of the determinant $\sqrt{-g}$ and on the inverse components $g^{\alpha\beta}$ within the trace term.
{Evaluating the remaining terms via the product rule yields two distinct structural contributions: one where the derivatives ($\partial_{\mu}$ and $\partial_{\nu}$) act on the inverse factors $g^{\rho\sigma}$ multiplying the second derivatives, and one where they act on the second derivatives themselves, generating third-order terms.}
After eliminating these third-order terms via the harmonic condition, the divergence $\mathcal{I}_\nu$ rigorously reduces to (see Appendix~\ref{app:divergence_I} for details):
\begin{align}
\mathcal{I}_\nu &= g^{\mu\alpha} (\partial_\mu g^{\rho\sigma}) \partial^2_{\rho\sigma} g_{\alpha\nu} - \frac{1}{2} g^{\alpha\beta} (\partial_\nu g^{\rho\sigma}) \partial^2_{\rho\sigma} g_{\alpha\beta} + g^{\mu\lambda} g^{\alpha\tau} (\Box_{\rm LB} g_{\lambda\tau}) \partial_\mu g_{\alpha\nu} \nonumber \\
&\quad - \frac{1}{2} g^{\mu\lambda} g^{\alpha\tau} (\Box_{\rm LB} g_{\lambda\tau}) \partial_\nu g_{\mu\alpha} - 2 g^{\rho\sigma} (\partial_\rho g^{\mu\alpha}) \partial^2_{\sigma\mu} g_{\alpha\nu} + g^{\rho\sigma} (\partial_\rho g^{\mu\alpha}) \partial^2_{\sigma\nu} g_{\mu\alpha}+Q_\nu.
\end{align}

The cubic residual vector $Q_{\nu} := {A}_{\nu} - {B}_{\nu}$ encodes all non-linear products of three first derivatives generated by the collision of derivatives acting on inverse metrics:
\begin{align}
{A}_{\nu} &= g^{\alpha\beta}g^{\sigma\kappa}g^{\rho\lambda}g^{\mu\tau}\left(\partial_{\alpha}g_{\rho\tau}\,\partial_{\beta}g_{\kappa\lambda}\,\partial_{\nu}g_{\sigma\mu}\right), \\
{B}_{\nu} &= 2g^{\alpha\beta}g^{\sigma\kappa}g^{\rho\lambda}g^{\mu\tau}\left(\partial_{\alpha}g_{\rho\tau}\,\partial_{\beta}g_{\kappa\lambda}\,\partial_{\mu}g_{\sigma\nu}\right). 
\end{align}

\subsection{Consistency and Vector Space Completeness}

The raised pseudo-tensor $U^\mu_\nu$ explicitly occupies the full 5-dimensional vector space spanned by the independent irreducible non-tensorial quadratic derivative combinations (First Fundamental Theorem of Invariant Theory \cite{Weyl1918,Weyl1939}):
\begin{align}\label{U's}
U^{(1)\mu}_\nu &:= g^{\mu\kappa}\partial_{\kappa}g^{\alpha\beta}\partial_{\alpha}g_{\nu\beta}, \nonumber\\
U^{(2)\mu}_\nu &:= g^{\mu\kappa}\partial_{\nu}g^{\alpha\beta}\partial_{\alpha}g_{\kappa\beta}, \nonumber\\
U^{(3)\mu}_\nu &:= g^{\mu\kappa}g^{\beta\gamma}g^{\alpha\epsilon}\partial_{\kappa}g_{\gamma\alpha}\partial_{\nu}g_{\beta\epsilon}, \nonumber\\
U^{(4)\mu}_\nu &:= g^{\mu\kappa}g^{\beta\gamma}g^{\alpha\epsilon}\partial_{\alpha}g_{\kappa\gamma}\partial_{\beta}g_{\nu\epsilon}, \nonumber\\
U^{(5)\mu}_\nu &:= g^{\mu\kappa}g^{\beta\gamma}g^{\alpha\epsilon}\partial_{\alpha}g_{\kappa\gamma}\partial_{\epsilon}g_{\nu\beta}.
\end{align}

When taking the covariant divergence 
\begin{equation}
\mathcal{D}_\nu:=\nabla_\mu (U^\mu_\nu - \frac{1}{2}\delta^\mu_\nu U^\alpha_\alpha),
\end{equation}
these precise coefficients generate the exact distribution of $(\partial g)(\partial^2 g)$ and $(\partial g)^3$ structures required to perfectly cancel the divergence of the component-wise d'Alembertian terms under the harmonic gauge condition, ensuring energy-momentum conservation $\nabla_\mu T^\mu_\nu = 0$ by construction.

To determine ${U}_{\mu\nu}$ constructively without invoking global geometric theorems or assuming Pseudo-Riemannian curvature \textit{a priori}, one expands ${U}_{\nu}^{\mu}$ in terms of the complete vector space spanned by all independent irreducible rank-two quadratic first-derivative 
combinations\footnote{By Weyl's fundamental theorem of invariant theory, the most general rank-two quadratic first-derivative pseudo-tensor contains 11 independent structures. However, as demonstrated in Appendix \ref{app:uniqueness_coefficients} (Remark \ref{remarkB1}), dynamical consistency uniquely forces the 6 reducible structures to vanish, leaving only a complete basis of 5 irreducible topologies. Therefore, the restriction to a 5-parameter space is not an ad-hoc ansatz, but a mathematically necessary consequence of local energy-momentum conservation.}:
\begin{equation}
{U}_{\nu}^{\mu} = c_1 U^{(1)\mu}_{\nu} + c_2 U^{(2)\mu}_{\nu} + c_3 U^{(3)\mu}_{\nu} + c_4 U^{(4)\mu}_{\nu} + c_5 U^{(5)\mu}_{\nu}.
\end{equation}

Evaluating the covariant divergence of this general quadratic expansion under the harmonic gauge condition generates a non-trivial vector of second-order derivative terms $(\partial{g})(\partial^2{g})$ and cubic first-derivative terms $(\partial {g})^3$. Requiring exact pointwise cancellation with the divergence of the d'Alembertian terms forces a uniquely determined, overconstrained system of linear algebraic equations for the coefficients. 

This algebraic system admits the unique solution (see Appendix \ref{app:uniqueness_coefficients}):
\begin{equation}
(c_1, c_2, c_3, c_4, c_5) = \left(1,\, 1,\, {1}/{2},\, 1,\, -1\right).
\end{equation}

Inserting these unique coefficients back into the expansion yields exactly the non-linear completion of the wave equation. Comparing this result with the trace-reversed Einstein field equations, one finds that $\Box_{\rm LB} g_{\mu\nu} + U_{\mu\nu} = -2R_{\mu\nu}$ (see Appendix \ref{app:ricci_reduction}). Thus, the full Ricci tensor in harmonic coordinates \cite{Haro2026c} 
\begin{equation}\label{eq:Ricci_components}
\begin{split}
    R_{\mu\nu} = & -\frac{1}{2} g^{\alpha\beta} \partial^2_{\alpha\beta} g_{\mu\nu} - \frac{1}{2} \left(\partial_{\mu}g^{\alpha\beta}\partial_{\alpha}g_{\nu\beta} + \partial_{\nu}g^{\alpha\beta}\partial_{\alpha}g_{\mu\beta}\right) \\
    & - \frac{1}{4} g^{\beta\gamma} g^{\alpha\epsilon} \partial_{\mu}g_{\gamma\alpha}\partial_{\nu}g_{\beta\epsilon} - \frac{1}{2} g^{\beta\gamma} g^{\alpha\epsilon} \partial_{\alpha}g_{\mu\gamma}\left(\partial_{\beta}g_{\nu\epsilon} - \partial_{\epsilon}g_{\nu\beta}\right),
\end{split}
\end{equation}
emerges constructively and inevitably from the strict enforcement of local energy-momentum conservation, without assuming \textit{a priori} any Pseudo-Riemannian geometric structure.

It is worth noting why determining these five coefficients requires the explicit index analysis of Appendix~\ref{app:uniqueness_coefficients} rather than the uniqueness theorems of Hermann Weyl \cite{Weyl1918}, Élie Cartan \cite{Cartan1922}, or David Lovelock \cite{Lovelock1971}. Those classic proofs assume from the outset that the field operator is a generally covariant tensor across arbitrary coordinate systems, which allows one to pass to Riemann normal coordinates ($\partial g = 0$) and bypass the quadratic $\partial g\,\partial g$ sector altogether. Here, by contrast, the ansatz $\Box_{\rm LB}g_{\mu\nu} + U_{\mu\nu}$ is formulated strictly within the harmonic gauge (\ref{harmonic}), where the individual blocks are non-tensorial and normal coordinates cannot be used without breaking the wave operator. A priori, restricting to harmonic coordinates could admit non-covariant combinations whose divergence vanishes only modulo
$\partial_\mu(\sqrt{-g}g^{\mu\nu})=0$; Appendix~\ref{app:uniqueness_coefficients} rules out any such gauge-restricted artifacts and proves that energy-momentum conservation uniquely synthesizes the covariant Ricci tensor.

\

\textbf{A final remark is in order.} Although algebraically more involved than in the unimodular case, the harmonic gauge also admits an ordinary divergence conservation law upon introducing the Landau--Lifshitz pseudotensor $t_{\mathrm{LL}}^{\mu\nu}$ \cite{Landau,Fock1964}:
\begin{equation}
    \partial_{\mu}\left((-g)\bigl(T^{\mu\nu} + t_{\mathrm{LL}}^{\mu\nu}\bigr)\right) = 0.
\end{equation}

In harmonic coordinates, its explicit components take the form \cite{Landau}
\begin{equation}
\begin{aligned}
(-g)t^{\mu\nu}_{\mathrm{LL}} := \frac{1}{16\pi G} \Bigg( & \frac{1}{2} g^{\mu\nu} g_{\lambda\alpha} \partial_\rho \mathfrak{g}^{\lambda\beta} \partial_\beta \mathfrak{g}^{\alpha\rho} - (g^{\mu\lambda} \partial_\rho \mathfrak{g}^{\beta\nu} + g^{\lambda\nu} \partial_\rho \mathfrak{g}^{\mu\beta}) g_{\alpha\beta} \partial_\lambda\mathfrak{g}^{\alpha\rho} \\
& + g_{\lambda\alpha} g^{\beta\rho} \partial_\beta \mathfrak{g}^{\mu\lambda} \partial_\rho \mathfrak{g}^{\alpha\nu} + \frac{1}{8} \left( 2 g^{\mu\lambda} g^{\alpha\nu} - g^{\mu\nu} g^{\lambda\alpha} \right) \left( 2 g_{\beta\rho} g_{\sigma\tau} - g_{\rho\sigma} g_{\beta\tau} \right) \partial_\lambda \mathfrak{g}^{\beta\tau} \partial_\alpha \mathfrak{g}^{\rho\sigma} \Bigg),
\end{aligned}
\end{equation}
where $\mathfrak{g}^{\alpha\beta} := \sqrt{-g}\,g^{\alpha\beta}$ denotes the ``gothic'' inverse  density, which satisfies the harmonic gauge condition $\partial_{\alpha}\mathfrak{g}^{\alpha\beta} = 0$.

\section{The Spin-$2$ Field Viewpoint}\label{sec:spin}

An alternative conceptual route to General Relativity emerged from particle physics in the mid-20th century, pioneered by Robert Kraichnan \cite{Kraichnan1955} and Suraj Gupta \cite{Gupta1954}, and most notably developed by Richard Feynman \cite{Feynman1995}, Steven Weinberg \cite{Weinberg1965}, and Stanley Deser \cite{Deser1970}. In this framework, gravitation is stripped of its primary geometric character and reinterpreted as a standard non-linear field theory operating on an \textit{a priori} flat Minkowski background $\eta_{\mu\nu}$.

The starting point of the field-theoretic approach is to postulate a massless, symmetric rank-$2$ tensor field $h_{\mu\nu}$ representing a particle of spin-$2$ (the graviton). At the linear level, the free propagation of $h_{\mu\nu}$ is described by the linear Fierz--Pauli action \cite{Pauli}, and its minimal coupling to matter is governed by the interaction term $\mathcal{L}_{\rm int} \sim h_{\mu\nu} T^{\mu\nu}$.

However, because the gravitational field itself carries energy and momentum, consistency demands that $h_{\mu\nu}$ must also couple to its own stress-energy. This requirement forces an iterative scheme:
\begin{equation}
\Box h_{\mu\nu} \sim T_{\mu\nu} + t_{\mu\nu}^{(1)}(h) + t_{\mu\nu}^{(2)}(h^2) + \dots
\end{equation}

Feynman and Weinberg argued---by enforcing consistency order by order in the lowest perturbative vertices \cite{Feynman1995} and through Lorentz invariance of the $S$-matrix \cite{Weinberg1965}---that continuing this iterative bootstrap indefinitely reproduces the formal weak-field expansion of General Relativity. However, within the  perturbation series around Minkowski spacetime, this iterative bootstrap remained a formal infinite expansion rather than yielding a closed-form second-order differential operator. Deser elegantly circumvented this endless iteration by shifting to a first-order (Palatini) formalism, demonstrating that the action closes at cubic order in a single step when the metric and the affine connection are treated as independent dynamical variables \cite{Deser1970}. By contrast, our approach achieves an exact, closed-form non-perturbative completion directly in the second-order metric formulation, without introducing independent connection variables or auxiliary variational structures.

From the viewpoint of Feynman and Weinberg, spacetime curvature is not a fundamental postulate, but rather an effective mathematical byproduct resulting from the universal coupling of a massless spin-$2$ field to energy-momentum. Despite its success, this reductionist program carries intrinsic epistemological and structural limitations:
\begin{enumerate}
\item \textbf{The Unobservable Background and the Wave Operator:} The flat Minkowski background $\eta_{\mu\nu}$ used to define the spin-$2$ field becomes physically unobservable, as all physical rods and clocks measure proper time governed dynamically by the full components of the invariant $g_{\mu\nu} = \eta_{\mu\nu} + h_{\mu\nu}$. Because gravity fundamentally modulates the proper time infinitesimal $ds$, the flat-space d'Alembertian ($\Box$) employed in the perturbative approach is physically inadequate. The natural differential operator governing wave dynamics must be the Laplace--Beltrami operator ($\Box_{\mathrm{LB}}$) natively associated with this dynamically modulated $ds$. Once this chronometric coupling is recognized, our constructive procedure provides the natural completion within the adopted second-order wave-operator framework: the exact non-linear field equations emerge in a single step upon adopting the physical wave operator and enforcing local energy-momentum conservation, completely bypassing the need for infinite flat-space iterations.

    \item \textbf{Perturbative vs. Non-Perturbative Emergence:} 
The spin-$2$ field-theoretic approach reconstructs General Relativity through the self-coupling of a massless spin-$2$ field around a flat Minkowski background. While this provides a consistent classical route to the theory, its perturbative quantization leads to the well-known problem of non-renormalizability. From our perspective, this difficulty also points to a deeper conceptual distinction: the microscopic and macroscopic descriptions are different maps of physical reality, each appropriate to its own level of description. Attempting to force the macroscopic gravitational continuum into a microscopic quantum-field-theoretic map risks confusing these distinct representational frameworks. The non-renormalizability of perturbative quantum gravity is a technical limitation of this particular microscopic formulation, not a failure of the classical spin-$2$ construction or of General Relativity itself.
\end{enumerate}

While Deser achieved formal closure by expanding the fundamental dynamical variables, our constructive approach achieves exact, non-perturbative closure directly within the standard second-order formulation. By systematically classifying all possible quadratic first-derivative contractions ($\partial g \partial g$) into Weyl's irreducible basis topologies $\{U^{(1)}, \dots, U^{(5)}\}$---paralleling the invariant structures utilized in the action formalisms of Gupta and Feynman, but executed here directly on the chronometric equations of motion---we demonstrate that strict differential consistency ($\nabla_\mu T^{\mu\nu} = 0$) in harmonic gauge forces these self-interaction blocks to assemble uniquely in a single step. Consequently, Pseudo-Riemannian geometry emerges neither as an \textit{a priori} postulate nor as a perturbative spin-$2$ artifact, but as the exact, macroscopic realization of dynamical consistency.

\section{Remarks on Natural Philosophy: \textit{Conatus} and Proper Time}
\label{sec:natural_philosophy}

Beyond its technical derivation, the bottom-up framework developed in this work invites a reassessment of several foundational concepts of natural philosophy, beginning with the tension between passive \textit{inertia} and active \textit{conatus}. When Johannes Kepler introduced the term \textit{inertia} into physics \cite{Kepler1618}, he drew directly on its Latin root \textit{iners} (``sluggish'' or ``idle'') to denote a purely passive property of matter: its resistance to being moved and its natural tendency to remain at rest whenever an impelling agent ceases to act. A decisive conceptual shift occurred when Ren\'e Descartes, in his \textit{Principia Philosophiae} (1644) \cite{Descartes1644}, introduced the hypothesis that uniform rectilinear motion and rest share the exact same ontological status as effortless ``natural'' states, reinterpreting persistence in motion as \textit{conatus}. Yet Descartes's purely relational view of space left an unresolved ambiguity---uniform motion relative to what?---which Isaac Newton resolved by postulating an immutable, unobservable \textit{absolute space} and \textit{absolute time}. Building on Robert Hooke's mechanical picture of orbital motion---first outlined in his 1674 treatise \cite{Hooke1674} and sharpened in his celebrated 1679--1680 correspondence with Newton \cite{HookeNewton1679}, where Hooke compounded a tangential rectilinear tendency with an inward central deflection following an inverse-square law---Newton codified this synthesis in the first edition of the \textit{Philosophiae Naturalis Principia Mathematica} \cite{Newton1687}. He did so by sharply separating the innate force of inertia (\textit{vis insita}, or paradoxically \textit{vis inertiae}, ``force of inactivity'') from gravity as an external impressed force (\textit{vis impressa}). Crucially, however, this Cartesian--Newtonian framework rests on an unphysical idealization: it defines pure rectilinear persistence in a hypothetical arena where gravity has been switched off. Yet gravitation couples universally to all forms of energy and cannot be shielded or turned off; a global, gravity-free inertial frame does not exist in nature.

Furthermore, Newton's separation of inertia and gravity left the ontological nature of gravitation deliberately unresolved. Confronted by Christiaan Huygens and the Cartesians \cite{Huygens1690}, who objected that gravitational attraction across empty space revived unintelligible ``occult qualities,'' Newton famously replied in the \textit{General Scholium} of the second edition with his maxim \textit{Hypotheses non fingo} (``I feign no hypotheses'') \cite{Newton1713}. Refusing to invent mechanical fictions akin to Cartesian vortices to explain what gravity \textit{is} ontologically, Newton restricted himself to describing mathematically \textit{how} it acts. In modern terms, he crafted an extraordinarily accurate mathematical map of gravitational accelerations while remaining agnostic about the underlying physical territory---leaving gravity as an unexplained, instantaneous action at a distance.

Standard textbook General Relativity replaces the classical map of force vectors with a far more sophisticated geometric map, constructed from the intrinsic tensor fields of a four-dimensional Pseudo-Riemannian manifold. Notably, the original architects of these frameworks were acutely aware of the boundary between mathematical representation and physical ontology. Throughout the \textit{Principia}---most explicitly in Definition~VIII and the General Scholium\footnote{Echoing the familiar dictum about literary classics attributed to Mark Twain, Newton's \textit{Principia} has often been praised more widely than it has been read directly. As historians of mechanics such as Clifford Truesdell \cite{Truesdell1968} and I. Bernard Cohen \cite{Cohen1980} have shown, modern reception of the work has been heavily mediated through later analytical and textbook reformulations, which do not always preserve the methodological subtleties and qualifications of Newton's original approach.}---Newton emphasized that he treated forces strictly mathematically, cautioning readers against interpreting ``attractions'' or ``impulses'' as ultimate physical causes; it was subsequent generations who reified those vectors into ontological realities acting across absolute space. Similarly, Bernhard Riemann himself, in his foundational 1854 \textit{Habilitationsvortrag} \cite{Riemann1867}, stressed that the metric relations of a continuous manifold cannot be deduced from pure geometry alone, but must ultimately originate in the physical interactions binding it together. It was rather the later G\"ottingen mathematical school---building on Hermann Minkowski's four-dimensional spacetime and spearheaded by Felix Klein, David Hilbert (who sought to axiomatize all of physics from geometric first principles), and Hermann Weyl---that inverted this order into a systematic geometrization of physics \cite{Hilbert1916,Klein1917}. Albert Einstein repeatedly resisted reducing gravity to pure geometry, insisting instead on the physical reality of the gravitational field---treating the $g_{\mu\nu}$ components fundamentally as dynamical potentials and the affine connection as the field strength \cite{Einstein1928,Lehmkuhl2014}. Yet, modern textbook expositions largely inherited the G\"ottingen perspective, often reifying the curved spacetime manifold as an \textit{a priori} physical arena. As Alfred Korzybski famously cautioned in \textit{Science and Sanity} (1933), ``the map is not the territory'' \cite{Korzybski1933}.

Within our chronometric view, separating the map from the territory restores both Newton's restraint and Riemann's physical priority: gravity is neither an instantaneous action at a distance nor a kinematic constraint imposed by an \textit{a priori} geometry, but the \textit{potentia} of surrounding matter---a local causal influence, propagated dynamically via the wave operator $\Box_{\rm LB}$, that alters the flow of proper time by deforming the flat Minkowski interval $ds_{\rm Min}^2$ into the dynamical proper-time invariant $ds^2 = g_{\mu\nu}dx^\mu dx^\nu$. Whereas Riemann originally conceived the metric in terms of spatial rods, relativity grounds it directly in physical clocks. This chronometric mechanism compels a fundamental paradigm shift in our conception of inertial motion, replacing Keplerian sluggishness and the kinematic Cartesian \textit{conatus} with the Stoic notion of \textit{horme}---the intrinsic striving of living beings to preserve themselves, translated into Latin as \textit{conatus} by Cicero---which Baruch Spinoza later elevated to a deep ontological principle in his \textit{Ethica} \cite{Spinoza1677}: the active endeavor of every thing to persevere in its own being (\textit{in suo esse perseverare}).

Physically, persevering in existence for a massive system is measured not by passive rest or rectilinear motion in a fictitious empty space, but by the maximization of its own elapsed proper time: as John Archibald Wheeler encapsulated in his \textit{Principle of Maximal Aging} \cite{TaylorWheeler1992}, a freely falling massive body follows the worldline that accumulates the greatest number of ticks of its local clock ($\delta \int ds = 0$), while massless radiation is constrained by the null boundary ($ds^2 = 0$) of the very same chronometric structure. Modern relativistic physics still refers to such timelike paths as ``inertial trajectories'' out of historical habit, yet their physical content is the exact opposite of Kepler's \textit{iners}: rather than sluggishly resisting change, a freely falling system actively locally  maximizes its own duration of existence.

{
Furthermore, as shown in Remark~\ref{rem:variational_euler}, this variational character extends seamlessly from pressureless matter to an interacting homentropic perfect fluid. For any non-vanishing specific enthalpy \(w=(\rho+p)/n\), the squared conformal factor \((w/m_0)^2>0\) in the effective potentials \(\widetilde{g}_{\mu\nu}=(w/m_0)^2g_{\mu\nu}\) preserves the Lorentzian signature of \(g_{\mu\nu}\). Moreover, for ordinary matter satisfying \(\rho+p>0\), one has \(w>0\), so that the conformally rescaled proper-time element \(d\widetilde{s}=(w/m_0)\,ds\) remains strictly positive along future-directed timelike worldlines. 
The fluid trajectories governed by \(\delta\int d\widetilde{s}=0\) therefore locally maximize the conformally 
 enthalpy-weighted
 duration \(\int d\widetilde{s}=m_0^{-1}\int w\,ds\) between fixed endpoints, within a sufficiently small neighbourhood. The chronometric \emph{conatus} of freely falling matter is thus preserved, but generalized: internal pressure does not oppose this perseverance through an external force; rather, it modulates from within the very measure of duration that the fluid locally maximizes.

As Einstein showed in 1911, by applying the Equivalence Principle, a gravitational field affects the rate of clocks and hence the proper time accumulated by matter, thereby modulating its chronometric \emph{conatus} \cite{Einstein1911}. The conformal formulation developed here raises a further possibility: if the gravitational potentials determine the rate at which proper time accumulates, could the specific enthalpy of matter, through its contribution to the effective gravitational potentials, also modify this rate and thus the chronometric \emph{conatus} itself? This remains a heuristic physical hypothesis, whose operational meaning and empirical validity call for further investigation.

}

At the same time, our bottom-up framework carries this anti-reification stance one step further than Einstein's own realism regarding autonomous physical fields, and stands in deliberate contrast to mathematical Platonism, which treats equations as inhabitants of an independent abstract realm. Adopting the strict nominalism championed by William of Ockham alongside an operational outlook, we align Newton's \textit{Hypotheses non fingo} and Korzybski's map--territory distinction with Niels Bohr's celebrated dictum that ``it is wrong to think that the task of physics is to find out how nature is; physics concerns what we can say about nature'' \cite{Petersen1963}. Accordingly, when we speak of mass-energy exerting a causal \textit{potentia} that modulates the \textit{conatus} of surrounding bodies, we make no metaphysical claim about the hidden ontological substance of spacetime or of a gravitational ``field.'' Coordinates, curvature tensors, and wave operators remain human descriptive instruments; the empirical content of the theory lies solely in specifying quantitatively how surrounding matter-energy governs the measurable local ticks of proper time ($ds^2$) along which physical systems persist.

This persistent impulse to reify mathematical mediators into autonomous substances reflects a deep-seated reluctance in physics to abandon the ancient \textit{horror vacui}---a recurring return of the aether in ever more abstract guises. From Aristotle's fifth element and the Cartesian--Huygensian plenum of vortices, the refusal to accept causal influence across empty space led James Clerk Maxwell first to model electromagnetism via mechanical vortices and idle wheels \cite{Maxwell1861}, and subsequently to endow the electromagnetic field itself with physical reality seated in a luminiferous aether. When Einstein dispensed with the mechanical aether in 1905, he still replaced Newton's absolute space with the rigid, immutable stage of Minkowski spacetime ($\eta_{\mu\nu}$), which acts upon matter without being acted upon. In General Relativity, this static theatre gave way to a dynamical ontology along two closely related paths: Einstein himself moved from describing space as endowed with physical qualities---a non-mechanical ``gravitational aether'' \cite{Einstein1920}---to elevating the continuous \emph{field} $g_{\mu\nu}$ into the primary physical reality, reducing space to a mere structural quality of the field \cite{Einstein1954,Lehmkuhl2014}; subsequently, the geometric tradition reified the curved spacetime manifold itself into a flexible fabric in which, as John Archibald Wheeler famously summarized, \textit{matter tells spacetime how to curve and spacetime tells matter how to move} \cite{MTW1973}.
Even certain substantivalist ontological readings of modern quantum field theory and cosmology perpetuate this lineage: by reifying the vacuum into a dynamical plenum of quantum harmonic oscillators and zero-point fluctuations \cite{Schrodinger1939} that interacts with matter and seeds primordial cosmological perturbations, the ``void'' is once again populated by an active physical medium.
By contrast, adhering at the classical level to the methodological restraint of Newton's \textit{General Scholium} and Bohr's operationalism, the chronometric perspective dispenses with any substantive ontological stage altogether: rather than asking \emph{what} the underlying territory is in itself, it confines the mathematical map strictly to describing \emph{how} the \textit{potentia} of matter modulates local clock rates ($ds^2$) and thereby governs the \textit{conatus} of surrounding bodies.

Finally, this chronometric perspective overturns what might be called \textit{Newtonian inertial supremacy}---the privileged epistemic hierarchy in which an idealized, gravity-free observer claims the exclusive right to dictate true kinematics from above, dismissing the dynamical effects experienced by all other observers as mere ``fictitious forces.'' Such an unobservable absolute frame had already drawn incisive criticism from Gottfried Wilhelm Leibniz in his correspondence with Samuel Clarke \cite{LeibnizClarke1717}, George Berkeley in \textit{De Motu} \cite{Berkeley1721}, Jean le Rond d'Alembert (whose dynamical balance principle bypassed privileged inertial frames) \cite{DAlembert1743}, and most forcefully Ernst Mach in \textit{The Science of Mechanics} \cite{Mach1883}, who rejected attributing inertial effects to acceleration relative to an empty, non-acting space.
{Einstein brought this relational critique to a striking physical reversal. In Newtonian mechanics, an observer approximately at rest with respect to the Earth's surface may be taken as an inertial observer, while an accelerating pilot is described from that standpoint as occupying a non-inertial frame, in which inertial forces appear as fictitious. In General Relativity, however, the roles are reversed in a fundamental sense: freely falling observers are locally inertial, whereas observers held at rest relative to the Earth's surface are accelerated. The very distinction between inertial and non-inertial motion thus depends on the physical framework adopted. Yet why should the observer on the ground, or in the control tower, enjoy any privileged right to dismiss the dynamical effects experienced by the pilot? Following the spirit of d'Alembert's principle, the equations of motion need not grant an inertial observer any exclusive authority over the description of physical reality. The pilot's experienced $g$-forces are genuine physical effects, just as are the changes in the reading of a weighing scale beneath a passenger in an accelerating elevator: the scale registers a greater weight when the elevator accelerates upwards and a smaller one when it accelerates downwards. These effects remain physically real even if they are represented as inertial forces in the tower observer's description.

}

In conclusion, it is crucial to emphasize that \textit{conatus} and \textit{potentia} are not introduced here merely as historical metaphors, but as the epistemological map through which the dynamics of gravitation can be consistently described. We do not attempt to formulate hypotheses concerning the ultimate physical essence of this \textit{potentia}. By shifting from the language of external forces and \textit{a priori} geometric manifolds to the language of modified proper time, we are not changing the territory of reality, but offering a dynamically consistent, bottom-up chart with which to navigate it.

\section{Conclusions}\label{sec:conclusions}

In this paper, we have presented a systematic constructive derivation of Einstein's field equations in harmonic coordinates, demonstrating how General Relativity can be built from dynamical principles rather than abstract geometric postulates. By tracing the evolution of proper time along particle worldlines, we first established how static and moving matter distributions alter the Minkowski line element. As noted in Section \ref{sec:linear_gravity}, while the uniform field is strictly derived from kinematics, its generalization to arbitrary potentials relies on a physically motivated conformastatic working hypothesis. Building upon this phenomenological seed, we naturally recovered the linearized Einstein equations without prior reliance on Pseudo-Riemannian curvature.

Our central mathematical result concerns the full non-linear completion of the linear wave equation obtained in the weak-field approach. By generalizing the d'Alembertian to the Laplace-Beltrami operator $\Box_{\rm LB}$ and expanding the non-linear self-interaction block $U_{\mu\nu}$ over the 5-dimensional vector space of irreducible quadratic first-derivative combinations $U^{(k)}_{\mu\nu}$, we evaluated the differential consistency condition enforced by local energy-momentum conservation, $\nabla_\mu T^{\mu}_{\nu} = 0$. The exact pointwise cancellation of third-order derivative terms ($\partial^3 g$) and cubic residuals $Q_\nu$ under the de Donder gauge yields a unique algebraic solution for the expansion coefficients within this specific invariant basis:
\begin{equation}
(c_1, c_2, c_3, c_4, c_5) = \left(1, 1, {1}/{2}, 1, -1\right).
\end{equation}

This specific combination exactly reconstructs the full Ricci tensor in harmonic coordinates as the unique minimal second-order completion within the specified quadratic first-derivative ansatz (\ref{nonlinear_wave_eq}). Because our constructive requirement of exact matter-energy conservation algebraically fixes these internal coefficients without relying on geometric postulates, the resulting dynamical synthesis provides a profound physical counterpart to Lovelock's celebrated theorem. In four dimensions, Lovelock starts from a completely unspecified second-order tensor operator and exploits general diffeomorphism covariance to evaluate it in Riemann normal coordinates at any event $P$: there, all first derivatives ($\partial g$) vanish identically and the surviving second derivatives ($\partial^2 g$) reduce directly to the Riemann curvature tensor, bypassing the quadratic $\partial g\,\partial g$ sector altogether. In our bottom-up program, by contrast, although the total field operator is a true tensor proportional to the trace-reversed stress-energy tensor, its explicit wave-theoretic form $\Box_{\rm LB}g_{\mu\nu} + U_{\mu\nu}$ holds specifically in harmonic coordinates. Passing to Riemann normal coordinates would violate the harmonic condition 
 (\ref{harmonic}), destroying the wave-operator structure and leaving no physical principle to determine the form of the tensor without postulating curvature \emph{a priori}. Staying within the harmonic gauge preserves this physical wave structure at the price of keeping $\partial g \neq 0$, making the explicit algebraic closure of the quadratic first-derivative sector in Appendix~\ref{app:uniqueness_coefficients} the necessary dynamical step to construct the tensor explicitly from wave consistency and energy-momentum conservation.

This constructive path offers key epistemological and physical insights:
\begin{enumerate}
    \item \textbf{Ontological Status of Geometry:} 
     Spacetime curvature need not be regarded as an \textit{a priori} physical entity or as an independently existing stage upon which dynamics takes place. General Relativity admits alternative geometrical formulations, such as teleparallel and symmetric teleparallel  \cite{Weitzenbock1923,Beltran}, which provide different mathematical representations of the same underlying physical dynamics. Among these, the standard Pseudo-Riemannian formulation stands out, in our view, for its exceptional mathematical elegance and unifying power. The bottom-up construction developed here shows that this geometrical framework need not be postulated from the outset: it arises naturally from physical principles and the requirements of dynamical consistency and local energy-momentum conservation. This emergence, however, should not be mistaken for an ontological claim: geometry remains a mathematical representation of physical reality, not reality itself. It is a map, not the territory.
    
        \item \textbf{An Alternative Route to Energy-Momentum Consistency:} Einstein's search for generally consistent gravitational field equations in 1915 involved a close interplay between the choice of field equations, the role of the gravitational energy-momentum pseudo-tensor $t^\mu_\nu$, and restrictions on admissible coordinate systems, particularly unimodular coordinates. The present approach adopts a different methodological starting point: rather than introducing the gravitational field operator geometrically and subsequently examining its compatibility with energy-momentum conservation, we take exact local matter energy-momentum conservation as a fundamental dynamical requirement. This shifts the logical direction of the construction, allowing the consistency of the matter dynamics to guide the determination of the gravitational field operator, without requiring curvature tensors or gravitational energy-momentum pseudo-tensors as foundational ingredients.

    \item \textbf{Contrast with the Spin-2 Program:} The standard spin-2 field-theoretic approach, as developed by Feynman and Weinberg, constructs General Relativity through an iterative self-coupling procedure around a flat Minkowski background. Deser subsequently demonstrated that this construction admits a closed formulation at cubic order within a first-order (Palatini) framework. Our approach provides a complementary route, achieving an exact non-perturbative closure directly in the standard second-order formulation, uniquely within the harmonic gauge and the chosen invariant basis. Rather than building the gravitational self-interaction through successive perturbative iterations, we determine its complete quadratic first-derivative structure in a single step, by enforcing the exact differential consistency condition associated with local energy-momentum conservation.
    \end{enumerate}

By decoupling the foundational principles of General Relativity from strictly requiring \textit{a priori} geometric axioms, this framework provides a clear conceptual bridge between classical continuum dynamics and field-theoretic formulations of gravity, reaffirming that the non-linearity of Einstein's theory is fundamentally the statement that gravitational energy gravitates.

In closing, this work is also intended as a tribute to Albert Einstein and to the remarkable eight-year intellectual journey that led him, largely through his own physical insight, to the formulation of General Relativity. Beginning in 1907 with his first formulation of the Principle of Equivalence \cite{Einstein1907}, this journey proceeded through his early, unsuccessful attempts to incorporate gravitation into Special Relativity, eventually leading around the end of 1912 to what may be called his \textit{geometrical turn}: the decisive encounter with the absolute differential calculus, through which the mathematical framework needed to complete the theory gradually took shape with the crucial collaboration of Marcel Grossmann. This long process culminated on 25 November 1915, when Einstein presented the final gravitational field equations to the Prussian Academy of Sciences \cite{Einstein1915}.

With the greatest humility, we have attempted to retrace this intellectual path up to the point at which geometry became an essential part of Einstein's approach. In doing so, our perspective resonates---in a different yet complementary key---with a philosophical sensibility that Einstein himself deeply admired. Whereas Einstein famously embraced Spinoza's \textit{Deus sive Natura} as the expression of an immanent, self-contained rational harmony in Nature, our chronometric framework draws operationally on Spinoza's complementary notion of \textit{conatus}: the intrinsic tendency of physical systems to persevere in existence by maximizing their own proper time, without requiring an \textit{a priori} geometric stage to dictate their motion from outside. Taking a retrospective route from that 1912 crossroads, we have sought to continue the original physical reasoning by allowing the geometrical structure to emerge directly from dynamical consistency. The results suggest that the two paths ultimately converge: one arriving at General Relativity through the geometrical insight that completed Einstein's journey, the other recovering its geometrical structure from physical principles and local energy-momentum conservation. In this sense, the present work is not intended to replace Einstein's geometrical formulation, but rather to retrace its physical origins and show how naturally that remarkable mathematical structure emerges from them.

\begin{acknowledgments}
This work  is supported by the Spanish grant PID2025-168310NB-I00 funded by MICIU/AEI/10.13039/501100011033 and by FSE+.
\end{acknowledgments}

\appendix
\section{Explicit Evaluation of the Divergence $\mathcal{I}_\nu$}
\label{app:divergence_I}

In this Appendix, we present the step-by-step derivation of the covariant divergence of the linearized wave operator block:
\begin{equation}
\mathcal{I}_\nu := \nabla_\mu E_\nu^\mu = \nabla_\mu \left( g^{\mu\alpha}\Box_{\rm LB} g_{\alpha\nu} - \frac{1}{2}\delta^\mu_\nu g^{\alpha\beta}\Box_{\rm LB} g_{\alpha\beta} \right).
\end{equation}

\subsection{Harmonic Gauge Simplifications}
We work strictly within harmonic (de Donder) coordinates, defined by:
\begin{equation}
\partial_\mu \left( \sqrt{-g} g^{\mu\sigma} \right) = 0 \Longleftrightarrow \Gamma^\mu :=g^{\alpha\beta}\Gamma^\mu_{\alpha\beta} = 0.
\label{app:harmonic_cond}
\end{equation}

Under Eq.~(\ref{app:harmonic_cond}), the action of the Laplace-Beltrami operator on any  component $g_{\alpha\beta}$ reduces to the reduced scalar d'Alembertian:
\begin{equation}
\Box_{\rm LB} g_{\alpha\beta} = \frac{1}{\sqrt{-g}}\partial_\rho\left(\sqrt{-g}g^{\rho\sigma}\partial_\sigma g_{\alpha\beta}\right) = g^{\rho\sigma} \partial^2_{\rho\sigma} g_{\alpha\beta}.
\end{equation}

\subsection{Landau-Lifshitz Divergence Identity and Expansion}
To evaluate $\nabla_\mu E_\nu^\mu$ for a mixed pseudo-tensorial quantity, we employ the exact Landau-Lifshitz identity for covariant divergences:
\begin{equation}
\nabla_\mu E_\nu^\mu = \frac{1}{\sqrt{-g}} \partial_\mu \left( \sqrt{-g} E_\nu^\mu \right) - \frac{1}{2} (\partial_\nu g_{\rho\lambda}) E^{\rho\lambda}.
\label{app:landau_lifshitz}
\end{equation}

Substituting the explicit form of $E_\nu^\mu$ into the first term of Eq.~(\ref{app:landau_lifshitz}) yields:
\begin{align}
\frac{1}{\sqrt{-g}} \partial_\mu \left( \sqrt{-g} E_\nu^\mu \right) =
\underbrace{\frac{1}{\sqrt{-g}} \partial_\mu \left( \sqrt{-g} g^{\mu\alpha} \Box_{\rm LB} g_{\alpha\nu} \right)}_{\text{Term I}} - \underbrace{\frac{1}{2\sqrt{-g}} \partial_\mu \left( \sqrt{-g} \delta^\mu_\nu g^{\alpha\beta} \Box_{\rm LB} g_{\alpha\beta} \right)}_{\text{Term II}}.
\label{app:split_terms}
\end{align}

Applying the Leibniz product rule to \textbf{Term I} by grouping factors as $(\sqrt{-g} g^{\mu\alpha}) \cdot (\Box_{\rm LB} g_{\alpha\nu})$:
\begin{align}
\text{Term I} &= \frac{1}{\sqrt{-g}} \left[ \sqrt{-g} g^{\mu\alpha} \partial_\mu \left( \Box_{\rm LB} g_{\alpha\nu} \right) + \partial_\mu \left( \sqrt{-g} g^{\mu\alpha} \right) \Box_{\rm LB} g_{\alpha\nu} \right] \nonumber \\
&= g^{\mu\alpha} \partial_\mu \left( \Box_{\rm LB} g_{\alpha\nu} \right) + \frac{1}{\sqrt{-g}} \partial_\mu \left( \sqrt{-g} g^{\mu\alpha} \right) \Box_{\rm LB} g_{\alpha\nu}.
\label{app:term1_expanded}
\end{align}

By virtue of the harmonic gauge condition (\ref{app:harmonic_cond}), the second piece of Term I vanishes identically.

For \textbf{Term II}, the Kronecker delta $\delta^\mu_\nu$ converts $\partial_\mu$ into a simple gradient $\partial_\nu$:
\begin{equation}
\text{Term II} = \frac{1}{2} \partial_\nu \left( g^{\alpha\beta} \Box_{\rm LB} g_{\alpha\beta} \right) + \frac{1}{2\sqrt{-g}} \partial_\nu(\sqrt{-g}) g^{\alpha\beta} \Box_{\rm LB} g_{\alpha\beta}.
\label{app:term2_expanded}
\end{equation}

Substituting the explicit derivatives of the invariant components, we evaluate the complete Landau-Lifshitz divergence $\mathcal{I}_\nu \equiv \nabla_\mu E_\nu^\mu$ by combining Term I, Term II, and the affine connection term. 

The affine connection term from Eq.~(\ref{app:landau_lifshitz}) splits into two pieces:
\begin{equation}
-\frac{1}{2}(\partial_\nu g_{\rho\lambda}) E^{\rho\lambda} = -\frac{1}{2}(\partial_\nu g_{\rho\lambda}) g^{\rho\alpha} g^{\lambda\beta} \Box_{\rm LB} g_{\alpha\beta} + \frac{1}{4} (\partial_\nu g_{\rho\lambda}) g^{\rho\lambda} g^{\alpha\beta} \Box_{\rm LB} g_{\alpha\beta}.
\end{equation}

Remarkably, this affine term acts as an exact counterpart to Term II. Using the identity $\partial_\nu g^{\alpha\beta} = -g^{\alpha\rho}g^{\beta\lambda}\partial_\nu g_{\rho\lambda}$, its first piece exactly cancels the principal derivative expansion $\frac{1}{2}(\partial_\nu g^{\alpha\beta}) \Box_{\rm LB} g_{\alpha\beta}$ from Term II. Simultaneously, its second piece exactly cancels the gradient contribution $\frac{1}{2\sqrt{-g}}\partial_\nu(\sqrt{-g}) g^{\alpha\beta} \Box_{\rm LB} g_{\alpha\beta}$.

Consequently, the combination of $-\text{Term II}$ and the affine term drastically simplifies, leaving only two surviving terms from the expansion of Term II:
\begin{equation}
-\text{Term II} - \frac{1}{2}(\partial_\nu g_{\rho\lambda}) E^{\rho\lambda} = - \frac{1}{2} g^{\alpha\beta} (\partial_\nu g^{\rho\sigma}) \partial^2_{\rho\sigma} g_{\alpha\beta} - \frac{1}{2} g^{\alpha\beta} g^{\rho\sigma} \partial^3_{\nu\rho\sigma} g_{\alpha\beta}.
\end{equation}

Adding the expansion of Term I to this result yields the complete, exact expression for the divergence:
\begin{align}
\mathcal{I}_\nu 
&= g^{\mu\alpha} (\partial_\mu g^{\rho\sigma}) \partial^2_{\rho\sigma} g_{\alpha\nu} - \frac{1}{2} g^{\alpha\beta} (\partial_\nu g^{\rho\sigma}) \partial^2_{\rho\sigma} g_{\alpha\beta} \nonumber \\
&\quad + g^{\mu\alpha} g^{\rho\sigma} \partial^3_{\mu\rho\sigma} g_{\alpha\nu} - \frac{1}{2} g^{\alpha\beta} g^{\rho\sigma} \partial^3_{\nu\rho\sigma} g_{\alpha\beta}.
\label{app:full_divergence}
\end{align}

\subsection{Transformation of Third-Order Derivatives}\label{sec:transformation_third_order}
Notice that all third-order derivatives are isolated in the second line of Eq.~(\ref{app:full_divergence}). Let us denote this third-order sector as $\mathcal{I}_\nu^{(3)} \equiv g^{\mu\alpha} g^{\rho\sigma} \partial^3_{\mu\rho\sigma} g_{\alpha\nu} - \frac{1}{2} g^{\alpha\beta} g^{\rho\sigma} \partial^3_{\nu\rho\sigma} g_{\alpha\beta}$. 

To eliminate these third-order terms, we differentiate the harmonic condition:
\begin{equation}
g^{\mu\alpha}\Gamma^\lambda_{\mu\alpha} = 0 \implies g^{\mu\alpha} \left( 2 \partial_\mu g_{\alpha\nu} - \partial_\nu g_{\mu\alpha} \right) = 0.
\label{app:gauge_constraint}
\end{equation}

Applying the Laplace-Beltrami operator $g^{\rho\sigma}\partial^2_{\rho\sigma}$ to Eq.~(\ref{app:gauge_constraint}) and distributing the derivatives yields the third-order terms along with a set of generated second-order terms $(\partial g)(\partial^2 g)$:
\begin{equation}
g^{\rho\sigma} g^{\mu\alpha} \left( 2 \partial^3_{\rho\sigma\mu} g_{\alpha\nu} - \partial^3_{\rho\sigma\nu} g_{\mu\alpha} \right) + 2 S_\nu = 0,
\end{equation}
where
\begin{equation}\label{S}
S_\nu = \frac{1}{2} g^{\rho\sigma} (\partial^2_{\rho\sigma} g^{\mu\alpha}) (2\partial_\mu g_{\alpha\nu} - \partial_\nu g_{\mu\alpha}) + g^{\rho\sigma} (\partial_\rho g^{\mu\alpha}) (2\partial^2_{\sigma\mu} g_{\alpha\nu} - \partial^2_{\sigma\nu} g_{\mu\alpha}).
\end{equation}

Dividing by $2$ isolates the exact differential identity for the third-order sector:
\begin{equation}
\mathcal{I}_\nu^{(3)} = g^{\mu\alpha} g^{\rho\sigma} \partial^3_{\mu\rho\sigma} g_{\alpha\nu} - \frac{1}{2} g^{\alpha\beta} g^{\rho\sigma} \partial^3_{\nu\rho\sigma} g_{\alpha\beta} = - S_\nu.
\label{app:3rd_cancel}
\end{equation}

Thus, the third-order derivatives do not simply vanish; they transform exactly into an additional set of second-order terms $-S_\nu$.

Therefore, substituting Eq.~(\ref{app:3rd_cancel}) into Eq.~(\ref{app:full_divergence}) yields
\begin{equation}
\mathcal{I}_\nu = g^{\mu\alpha} (\partial_\mu g^{\rho\sigma}) \partial^2_{\rho\sigma} g_{\alpha\nu} - \frac{1}{2} g^{\alpha\beta} (\partial_\nu g^{\rho\sigma}) \partial^2_{\rho\sigma} g_{\alpha\beta} - S_\nu,
\label{app:surviving_terms}
\end{equation}
with $S_\nu$ defined in Eq.~(\ref{S}).

Expanding $S_\nu$ via Eq.~(\ref{S}) and applying the exact inverse derivative identities
\begin{align}
\partial_\alpha g^{\mu\tau} &= - g^{\mu\rho} g^{\tau\lambda} \partial_\alpha g_{\rho\lambda}, \\
\partial^2_{\rho\sigma} g^{\mu\alpha} &= -g^{\mu\lambda} g^{\alpha\tau} \partial^2_{\rho\sigma} g_{\lambda\tau} + \left( g^{\mu\beta} g^{\lambda\kappa} g^{\alpha\tau} + g^{\mu\lambda} g^{\alpha\beta} g^{\tau\kappa} \right) \partial_\rho g_{\beta\kappa} \partial_\sigma g_{\lambda\tau},
\end{align}
separates $\mathcal{I}_\nu$ cleanly into its complete second-order $(\partial g)(\partial^2 g)$ sector and a purely cubic non-linear sector $Q_\nu$ of order $\mathcal{O}((\partial g)^3)$:
\begin{equation}
\mathcal{I}_\nu = \mathcal{I}_\nu^{(\text{2nd order})} + Q_\nu,
\label{app:I_2nd}
\end{equation}
where
{\begin{align}
\mathcal{I}_\nu^{(\text{2nd order})} &:= g^{\mu\alpha} (\partial_\mu g^{\rho\sigma}) \partial^2_{\rho\sigma} g_{\alpha\nu} - \frac{1}{2} g^{\alpha\beta} (\partial_\nu g^{\rho\sigma}) \partial^2_{\rho\sigma} g_{\alpha\beta} + g^{\mu\lambda} g^{\alpha\tau} (\Box_{\rm LB} g_{\lambda\tau}) \partial_\mu g_{\alpha\nu} \nonumber \\
&\quad - \frac{1}{2} g^{\mu\lambda} g^{\alpha\tau} (\Box_{\rm LB} g_{\lambda\tau}) \partial_\nu g_{\mu\alpha} - 2 g^{\rho\sigma} (\partial_\rho g^{\mu\alpha}) \partial^2_{\sigma\mu} g_{\alpha\nu} + g^{\rho\sigma} (\partial_\rho g^{\mu\alpha}) \partial^2_{\sigma\nu} g_{\mu\alpha},
\label{app:I_2nd_explicit}
\end{align}
and all cubic products of first derivatives $(\partial g)^3$ assemble uniquely into:}
\begin{equation}
Q_\nu := A_\nu - B_\nu,
\label{app:Q_def}
\end{equation}
where the non-linear contraction vectors $A_\nu$ and $B_\nu$ simplify (by dummy-index symmetry under $(\alpha,\rho,\tau) \leftrightarrow (\beta,\kappa,\lambda)$) to:
\begin{align}
A_\nu &= g^{\alpha\beta} g^{\sigma\kappa} g^{\rho\lambda} g^{\mu\tau} \left( \partial_\alpha g_{\rho\tau} \partial_\beta g_{\kappa\lambda} \partial_\nu g_{\sigma\mu} \right), \label{app:A_def} \\
B_\nu &= 2 g^{\alpha\beta} g^{\sigma\kappa} g^{\rho\lambda} g^{\mu\tau} \left( \partial_\alpha g_{\rho\tau} \partial_\beta g_{\kappa\lambda} \partial_\mu g_{\sigma\nu} \right). \label{app:B_def}
\end{align}

\section{Uniqueness of the Self-Interaction Coefficients $(c_1, \dots, c_5)$}
\label{app:uniqueness_coefficients}

In this Appendix, we prove that the requirement of differential consistency $\mathcal{I}_\nu + \mathcal{D}_\nu = 0$ uniquely determines the five coefficients of the self-interaction tensor $U^\mu_\nu$ to be $(c_1, c_2, c_3, c_4, c_5) = \left(1, 1, {1}/{2}, 1, -1\right)$.

\subsection{Completeness and Independence of the Quadratic Basis}

We consider all possible quadratic structures formed by two first derivatives of the metric components, schematically denoted by $\partial g \partial g$, carrying two free indices $U^\mu_\nu$. Specifically, the rank-$2$ pseudo-tensor $U_{\lambda\nu} \equiv g_{\lambda\mu} U^\mu_\nu$ can be systematically constructed by contracting two inverse metric tensors $g^{\sigma\gamma}$ with four of the six covariant indices of $\partial_\alpha g_{\beta\gamma}\,\partial_\delta g_{\epsilon\zeta}$, leaving two free indices $(\lambda, \nu)$. 

Counting the allowed inter-derivative pairings, choosing two indices out of three in each derivative factor yields $\binom{3}{2}\times\binom{3}{2} = 9$ index-pair combinations, which admit $2$ contraction wirings and $2$ assignments of the free indices $(\lambda,\nu)$, giving $36$ ordered contractions in total.

For compactness, some of these contractions naturally rewrite in terms of derivatives of the inverse tensor $\partial g^{\alpha\beta}$. In our classification, we restrict attention to irreducible inter-derivative contractions in which every inverse tensor $g^{\sigma\gamma}$ bridges the two derivative factors.
Internal contractions within a single derivative factor---namely the trace $g^{\sigma\gamma}\partial_\alpha g_{\sigma\gamma} = 2\partial_\alpha \ln\sqrt{-g}$ and the divergence $g^{\sigma\gamma}\partial_\sigma g_{\gamma\alpha}$, which in harmonic coordinates ($\Gamma^\alpha= 0$) likewise reduces to $\partial_\alpha \ln\sqrt{-g}$---factorize into lower-rank products of the determinant gradient $v_\alpha := g^{\rho\sigma}\partial_\alpha g_{\rho\sigma} = 2\partial_\alpha \ln\sqrt{-g}$.
As proven rigorously in Remark~\ref{remarkB1}, including this reducible sector in the general ansatz leaves the algebraic equations determining $(c_1,\dots,c_5)$ unaffected and uniquely forces all reducible coefficients to vanish by back-substitution, allowing us to focus first on the purely irreducible sector.

Within the class of quadratic first-derivative structures considered here, the invariant contractions under the local orthogonal group $O(3,1)$ can be classified using Weyl's fundamental theorem of invariant theory. Taking into account the symmetries of the metric tensor ($g_{\mu\nu} = g_{\nu\mu}$), the commutativity of the two derivative factors, and the freedom to relabel dummy indices, it is straightforward to verify algebraically that the $36$ initial contractions reduce to exactly five linearly independent index-contraction topologies, which are governed by the origin of the two uncontracted indices $(\lambda,\nu)$:
\begin{enumerate}
    \item Both free indices are derivative indices $(\alpha, \delta)$, leaving the symmetric pairs $(\beta,\gamma)$ and $(\epsilon,\zeta)$ to be contracted with each other. This yields a single symmetric structure:
    \begin{equation}
    U^{(3)}_{\lambda\nu} := g^{\beta\gamma} g^{\alpha\epsilon} \partial_\lambda g_{\gamma\alpha} \partial_\nu g_{\beta\epsilon}.
    \end{equation}
    \item One free index is a derivative index and the other is a component index, leaving one symmetric pair $(\beta,\gamma)$ to contract with $(\delta,\epsilon)$. This yields a single pair related by transposition ($U^{(1)}_{\nu\lambda} = U^{(2)}_{\lambda\nu}$):
    \begin{equation}
    U^{(1)}_{\lambda\nu} := \partial_\lambda g^{\alpha\beta} \partial_\alpha g_{\nu\beta}, \qquad U^{(2)}_{\lambda\nu} := \partial_\nu g^{\alpha\beta} \partial_\alpha g_{\lambda\beta}.
    \end{equation}
    \item Both free indices are component indices ($\gamma, \zeta$), leaving $(\alpha,\beta)$ and $(\delta,\epsilon)$ to be contracted. Because derivative and component indices are not symmetric under exchange, this yields two distinct symmetric contractions depending on whether the derivative indices contract with each other or cross-contract with the component indices:
    \begin{equation}
    U^{(4)}_{\lambda\nu} := g^{\beta\gamma} g^{\alpha\epsilon} \partial_\alpha g_{\lambda\gamma} \partial_\beta g_{\nu\epsilon}, \qquad U^{(5)}_{\lambda\nu} := g^{\beta\gamma} g^{\alpha\epsilon} \partial_\alpha g_{\lambda\gamma} \partial_\epsilon g_{\nu\beta}.
    \end{equation}
\end{enumerate}

Raising the first index via $U^{(k)\mu}_\nu = g^{\mu\lambda} U^{(k)}_{\lambda\nu}$ (and relabelling $\lambda \to \kappa$) immediately yields the canonical mixed basis (\ref{U's}).

To prove explicitly that $\{U^{(1)}, \dots, U^{(5)}\}$ are linearly independent off-shell, consider a general linear combination $\sum_{k=1}^5 a_k U^{(k)\mu}_\nu = 0$ evaluated at an arbitrary spacetime point $p$ where $g_{\mu\nu}(p) = \eta_{\mu\nu}$. At $p$, the first derivatives $d_{\alpha\beta\gamma} := \partial_\alpha g_{\beta\gamma}(p)$ (symmetric in $\beta,\gamma$) act as algebraically independent variables, and $\partial_\kappa g^{\alpha\beta}(p) = -\eta^{\alpha\rho}\eta^{\beta\sigma} d_{\kappa\rho\sigma}$. Working within the spatial sector  (where $\eta^{11}=\eta^{22}=\eta^{33}=-1$), we can isolate each coefficient sequentially:
\begin{itemize}
    \item \textit{Antisymmetric sector ($a_1 = a_2$):} Lowering the first index, $U^{(3)}_{\lambda\nu}, U^{(4)}_{\lambda\nu}, U^{(5)}_{\lambda\nu}$ are symmetric in $(\lambda,\nu)$, whereas $U^{(2)}_{\lambda\nu} = U^{(1)}_{\nu\lambda}$. Choosing $d_{133} = d_{323} = d_{332} = 1$ (all other $d_{\alpha\beta\gamma}=0$) yields $U^{(1)}_{12} = -(\eta^{33})^2 d_{133}d_{323} = -1$ and $U^{(1)}_{21} = 0$. Evaluating $\sum_{k=1}^5 a_k(U^{(k)1}_2 - U^{(k)2}_1) = 0$ gives $-(a_1 - a_2)\eta^{11} = 0$, hence $a_1 = a_2$.
    \item \textit{Isolation of $a_3$:} Setting only $d_{122} = 1$ and evaluating the $(\mu,\nu)=(1,1)$ component gives $U^{(3)1}_1 = \eta^{11}(\eta^{22})^2 (d_{122})^2 = \eta^{11} \neq 0$, while $U^{(1,2,4,5)1}_1 = 0$, forcing $a_3 = 0$.
    \item \textit{Isolation of $a_5$:} Setting only $d_{213} = d_{231} = 1$ yields $U^{(1,2,3)1}_1 = 0$. For $U^{(4)1}_1$, the only non-zero first factor requires $\alpha=2,\gamma=3$, which fixes $\beta=3,\epsilon=2$ and multiplies $d_{312}=0$, so $U^{(4)1}_1 = 0$. Conversely, $U^{(5)1}_1 = \eta^{11}\eta^{33}\eta^{22}(d_{213})^2 = \eta^{11} \neq 0$, forcing $a_5 = 0$.
    \item \textit{Isolation of $a_4$:} Setting only $d_{213} = d_{231} = 1$ and $d_{312} = d_{321} = 1$ (with $a_5=0$ already established) maintains $U^{(1,2,3)1}_1 = 0$ while giving $U^{(4)1}_1 = 2\eta^{11}\eta^{22}\eta^{33} d_{213}d_{312} = 2\eta^{11} \neq 0$, forcing $a_4 = 0$.
    \item \textit{Isolation of $a_1, a_2$:} Finally, with $a_3=a_4=a_5=0$ and $a_1=a_2$, setting only $d_{122} = 1$ and $d_{212} = d_{221} = 1$ yields $U^{(1)1}_1 = U^{(2)1}_1 = -\eta^{11}(\eta^{22})^2 d_{122}d_{212} = -\eta^{11} \neq 0$, giving $-2a_1\eta^{11} = 0$, so $a_1 = a_2 = 0$.
\end{itemize}

Combining this strict linear independence with the preceding combinatorial classification establishes that  the set $\{U^{(1)}, \dots, U^{(5)}\}$ forms a complete, five-dimensional basis for irreducible second-rank quadratic first-derivative structures. Consequently, the most general pseudo-tensor in this class takes the form:
\begin{equation}
U^{\mu}_{\nu}=c_1U^{(1)\mu}_\nu+c_2U^{(2)\mu}_\nu+c_3U^{(3)\mu}_\nu+c_4U^{(4)\mu}_\nu+c_5U^{(5)\mu}_\nu.
\end{equation}

\subsection{Explicit Divergence Contractions and Intermediate Derivations}
The master consistency equation requires $\mathcal{D}_\nu = -\mathcal{I}_\nu$, where $\mathcal{D}_\nu := \nabla_\mu \left( U^\mu_\nu - \frac{1}{2}\delta^\mu_\nu U^\alpha_\alpha \right)$. Since $\nabla_\mu \delta^\mu_\nu = 0$ and $\nabla_\nu U = \partial_\nu U$ for the trace $U := U^\alpha_\alpha$, this divergence splits as:
\begin{equation}
\mathcal{D}_\nu = \nabla_\mu U^\mu_\nu - \frac{1}{2}\partial_\nu U.
\end{equation}

Because both $\Box_{\mathrm{LB}}g_{\mu\nu}$ and the matter source $T_{\mu\nu}-\frac{1}{2}g_{\mu\nu}T$ in Eq.~(\ref{nonlinear_wave_eq}) are symmetric in $(\mu,\nu)$, the total self-interaction block $U_{\mu\nu}$ must likewise be symmetric ($U_{\mu\nu}=U_{\nu\mu}$). Since $U^{(3)}_{\kappa\nu}$, $U^{(4)}_{\kappa\nu}$, and $U^{(5)}_{\kappa\nu}$ are individually symmetric whereas $U^{(1)}_{\nu\kappa}=U^{(2)}_{\kappa\nu}$, symmetry of the total combination requires $c_1=c_2$. Applying the Landau--Lifshitz divergence identity (cf.\ Eq.~(\ref{app:landau_lifshitz})) to the \emph{total} pseudo-tensor $U^\mu_\nu = g^{\mu\kappa}U_{\kappa\nu}$ under the harmonic gauge condition $\partial_\mu(\sqrt{-g}g^{\mu\kappa})=0$ then yields:
\begin{align}
\nabla_\mu U^\mu_\nu 
&= \frac{1}{\sqrt{-g}}\partial_\mu\left(\sqrt{-g} g^{\mu\kappa} U_{\kappa\nu}\right) - \frac{1}{2}(\partial_\nu g_{\rho\lambda}) U^{\rho\lambda} \nonumber\\&
= g^{\mu\kappa} \partial_\mu U_{\kappa\nu} + \frac{1}{2}(\partial_\nu g^{\kappa\lambda}) U_{\kappa\lambda},
\end{align}
where we used $-\frac{1}{2}(\partial_\nu g_{\rho\lambda}) g^{\rho\kappa} g^{\lambda\sigma} = +\frac{1}{2}\partial_\nu g^{\kappa\sigma}$. Subtracting $\frac{1}{2}\partial_\nu(g^{\kappa\lambda} U_{\kappa\lambda}) = \frac{1}{2}(\partial_\nu g^{\kappa\lambda}) U_{\kappa\lambda} + \frac{1}{2}g^{\kappa\lambda}\partial_\nu U_{\kappa\lambda}$ produces an exact cancellation of the $\partial_\nu g^{\kappa\lambda}$ terms, reducing the divergence to the compact form:
\begin{equation}
\mathcal{D}_\nu = g^{\mu\kappa} \partial_\mu U_{\kappa\nu} - \frac{1}{2} g^{\kappa\lambda} \partial_\nu U_{\kappa\lambda}.
\label{app:D_compact}
\end{equation}

Because Eq.~(\ref{app:D_compact}) is manifestly linear in $U_{\kappa\nu}$, we expand $\mathcal{D}_\nu = \sum_{k=1}^5 c_k \mathcal{D}_\nu^{(k)}$. Moreover, since the first and second metric derivatives $(\partial g, \partial^2 g)$ act as independent jet variables at any spacetime point, the identity $\mathcal{D}_\nu + \mathcal{I}_\nu = 0$ decomposes cleanly into two independent differential orders:
\begin{equation}
\left(\sum_{k=1}^5 c_k \mathcal{D}_\nu^{(k)}\Big|_{\partial g\,\partial^2 g} + \mathcal{I}_\nu\Big|_{\partial g\,\partial^2 g}\right) + \left(\sum_{k=1}^5 c_k \mathcal{D}_\nu^{(k)}\Big|_{(\partial g)^3} + \mathcal{I}_\nu\Big|_{(\partial g)^3}\right) = 0.
\label{app:order_split}
\end{equation}

We first solve the linear second-order sector $\partial g\,\partial^2 g$ in full to determine $(c_1,\dots,c_5)$ uniquely, and then verify the exact closure of the cubic sector $(\partial g)^3$.

\subsubsection{Basis of Second-Order Contractions ($\partial g\,\partial^2 g$)}
\label{subsec:sec_order_basis}

Expanding the derivatives in Eq.~(\ref{app:D_compact}) and in $\mathcal{I}_\nu$ generates eight distinct second-order contraction structures carrying one free covariant index $\nu$. We group them into four natural families according to the position of $\nu$ and the trace contractions:
\begin{enumerate}
    \item \textit{Direct wave-operator contraction:}

\begin{equation}
W_{1,\nu} := - g^{\mu\kappa} g^{\alpha\gamma} g^{\beta\epsilon} (\partial_\alpha g_{\nu\beta}) \partial^2_{\mu\kappa} g_{\gamma\epsilon}.
\label{app:W1_def}
\end{equation}

    \item \textit{Trace-gradient wave contractions:}
    \begin{align}
    \widetilde{W}_{2,\nu} &:= \frac{1}{2} g^{\kappa\lambda} (\partial_\nu g^{\alpha\beta}) \partial^2_{\alpha\beta} g_{\kappa\lambda} = -\frac{1}{2} g^{\kappa\lambda} g^{\alpha\gamma} g^{\beta\epsilon} (\partial_\nu g_{\gamma\epsilon}) \partial^2_{\alpha\beta} g_{\kappa\lambda}, \label{app:Wtilde2_def} \\
    W_{2,\nu} &:= \frac{1}{2}  (\partial_\nu g^{\sigma\gamma}) \Box_{\rm LB} g_{\sigma\gamma} = -\frac{1}{2}  g^{\sigma\rho} g^{\gamma\tau} (\partial_\nu g_{\rho\tau}) \Box_{\rm LB} g_{\sigma\gamma}, \label{app:W2_def} \\
    W_{3,\nu} &:= \frac{1}{2} g^{\mu\kappa} g^{\beta\gamma} g^{\alpha\epsilon} (\partial_\beta g_{\nu\epsilon}) \partial^2_{\alpha\gamma} g_{\mu\kappa}. \label{app:W3_def}
    \end{align}
    \item \textit{Mixed second-derivative contractions (with $\nu$ on $\partial^2 g_{\nu\bullet}$):}
    \begin{align}
    M_{1,\nu} &:= g^{\mu\kappa} g^{\beta\gamma} g^{\alpha\epsilon} (\partial_\alpha g_{\kappa\gamma}) \partial^2_{\mu\beta} g_{\nu\epsilon}, \label{app:M1_def} \\
    M_{2,\nu} &:= g^{\mu\kappa} g^{\beta\gamma} g^{\alpha\epsilon} (\partial_\alpha g_{\kappa\gamma}) \partial^2_{\mu\epsilon} g_{\nu\beta}. \label{app:M2_def}
    \end{align}
    \item \textit{External-gradient contractions (with $\nu$ as an outer derivative $\partial^2_{\nu\bullet} g$):}
    \begin{align}
    E_{1,\nu} &:= g^{\kappa\lambda} g^{\alpha\gamma} g^{\beta\epsilon} (\partial_\alpha g_{\lambda\beta}) \partial^2_{\nu\kappa} g_{\gamma\epsilon}, \label{app:E1_def} \\
    E_{2,\nu} &:= g^{\rho\sigma} g^{\mu\lambda} g^{\alpha\tau} (\partial_\rho g_{\lambda\tau}) \partial^2_{\nu\sigma} g_{\mu\alpha}. \label{app:E2_def}
    \end{align}
\end{enumerate}

To establish explicitly that the eight contractions (\ref{app:W1_def})--(\ref{app:E2_def}) are linearly independent modulo the harmonic coordinate condition, evaluate an arbitrary linear combination
\begin{equation}
\alpha_1 W_{1,\nu} + \tilde{\alpha}_2 \widetilde{W}_{2,\nu} + \alpha_2 W_{2,\nu} + \alpha_3 W_{3,\nu} + \beta_1 M_{1,\nu} + \beta_2 M_{2,\nu} + \gamma_1 E_{1,\nu} + \gamma_2 E_{2,\nu} = 0
\label{app:lin_indep_8}
\end{equation}
at an arbitrary spacetime point $p$. Applying a constant linear coordinate transformation $x'^\mu = A^\mu_{\ \nu} x^\nu$---which preserves the harmonic condition $\Box_{\rm LB} x'^\mu = A^\mu_{\ \nu}\Box_{\rm LB} x^\nu = 0$ identically---we may bring the gravitational potentials at $p$ to canonical Minkowskian values $g_{\mu\nu}(p) = \eta_{\mu\nu} = \mathrm{diag}(+1,-1,-1,-1)$ without restricting their first or second derivatives, and fix the free index $\nu = 0$. Denoting $d_{\alpha\beta\gamma} := \partial_\alpha g_{\beta\gamma}(p)$ (symmetric in $\beta,\gamma$) and $s_{\alpha\beta\gamma\delta} := \partial^2_{\alpha\beta} g_{\gamma\delta}(p)$ (symmetric in $\alpha,\beta$ and in $\gamma,\delta$), differentiating the harmonic condition $g^{\mu\kappa}(2\partial_\mu g_{\kappa\beta} - \partial_\beta g_{\mu\kappa}) = 0$ imposes at $p$:
\begin{equation}
2\eta^{\mu\kappa}d_{\mu\kappa\rho} - \eta^{\mu\kappa}d_{\rho\mu\kappa} = 0, \qquad
L(s)_{\alpha\beta} := 2\eta^{\mu\kappa}s_{\mu\alpha\kappa\beta} - \eta^{\mu\kappa}s_{\alpha\beta\mu\kappa} = R_{\alpha\beta}(d^2),
\label{app:jet_harm}
\end{equation}
where $R_{\alpha\beta}(d^2)$ denotes the quadratic gradient terms generated by differentiating the inverse potentials $g^{\mu\kappa}$. 
Because the linear map $L$ is surjective---indeed, restricting to the ten-dimensional subspace spanned by $(s_p)_{\alpha\alpha\alpha\alpha}$ ($\alpha\in\{0,1,2,3\}$) and $(s_p)_{\alpha\alpha\alpha\beta} = (s_p)_{\alpha\alpha\beta\alpha}$ ($0\le\alpha<\beta\le 3$) with all other independent components set to zero diagonalizes Eq.~(\ref{app:jet_harm}) explicitly into
\begin{equation}
(s_p)_{\alpha\alpha\alpha\alpha} = \eta^{\alpha\alpha} R_{\alpha\alpha}(d^2), \qquad
(s_p)_{\alpha\alpha\alpha\beta} = (s_p)_{\alpha\alpha\beta\alpha} = \frac{1}{2}\eta^{\alpha\alpha} R_{\alpha\beta}(d^2) \quad (\alpha < \beta,\;\text{no sum over }\alpha),
\end{equation}
providing a particular solution $s_p$ for every harmonic $1$-jet $d$---any homogeneous solution $s_0$ satisfying $L(s_0) = 0$ yields another admissible $2$-jet $s_p + s_0$ solving Eq.~(\ref{app:jet_harm}). Since the left-hand side of Eq.~(\ref{app:lin_indep_8}) is linear in $s$, subtracting its evaluation at $(d, s_p)$ from that at $(d, s_p + s_0)$ implies that Eq.~(\ref{app:lin_indep_8}) must vanish on every pair $(d, s_0)$ with $d$ harmonic and $L(s_0) = 0$.
Choosing the eight admissible test configurations $(d, s_0)$ listed in Table~\ref{tab:lin_indep_8}---each satisfying $L(s_0)=0$ and vanishing on seven of the eight structures---isolates every coefficient in Eq.~(\ref{app:lin_indep_8}) individually ($\alpha_1 = \tilde{\alpha}_2 = \alpha_2 = \alpha_3 = \beta_1 = \beta_2 = \gamma_1 = \gamma_2 = 0$), proving that (\ref{app:W1_def})--(\ref{app:E2_def}) form a linearly independent basis of the eight-dimensional second-order subspace modulo the harmonic gauge.

{

\begin{table}[h!]
\centering
\caption{Admissible harmonic jet configurations $(d, s_0)$ with $L(s_0)=0$ at $g_{\mu\nu}(p)=\eta_{\mu\nu}$ isolating each of the eight second-order contractions (\ref{app:W1_def})--(\ref{app:E2_def}) for $\nu=0$ (unlisted independent components vanish; all remaining components follow from index symmetries).}
\label{tab:lin_indep_8}
\renewcommand{\arraystretch}{1.25}
\begin{tabular}{l l l c}
\hline\hline
\textbf{Isolated Term} & \textbf{Non-zero $d_{\alpha\beta\gamma}$} & \textbf{Non-zero $s_{\alpha\beta\gamma\delta}$ ($L(s_0)=0$)} & \textbf{Non-zero Value} \\
\hline
$W_{2,0}$ & $d_{012}=1$ & $s_{3312}=1$ & $W_{2,0} = +1$ \\
$\widetilde{W}_{2,0}$ & $d_{012}=1$ & $s_{1233}=1,\; s_{1323}=s_{2313}=\tfrac{1}{2}$ & $\widetilde{W}_{2,0} = +1$ \\
$W_{1,0}$ & $d_{102}=1$ & $s_{3312}=1$ & $W_{1,0} = +1$ \\
$W_{3,0}$ & $d_{102}=1$ & $s_{1233}=1,\; s_{1323}=s_{2313}=\tfrac{1}{2}$ & $W_{3,0} = -\tfrac{1}{2}$ \\
$M_{1,0}$ & $d_{123}=1$ & $s_{2301}=1$ & $M_{1,0} = -2$ \\
$M_{2,0}$ & $d_{123}=1$ & $s_{1203}=1$ & $M_{2,0} = -1$ \\
$E_{1,0}$ & $d_{123}=1$ & $s_{0213}=1$ & $E_{1,0} = -1$ \\
$E_{2,0}$ & $d_{123}=1$ & $s_{0123}=1$ & $E_{2,0} = -2$ \\
\hline\hline
\end{tabular}
\end{table}

}

\subsubsection{Full Second-Order Expansion and Linear System for $(c_1, \dots, c_5)$}\label{subsec:sec_order_system}
We now apply the Leibniz rule to $\mathcal{D}_\nu^{(k)} = g^{\mu\kappa}\partial_\mu U^{(k)}_{\kappa\nu} - \frac{1}{2}g^{\kappa\lambda}\partial_\nu U^{(k)}_{\kappa\lambda}$ for each basis tensor in Eqs.~(\ref{U's}),
keeping all terms of order $\partial g\,\partial^2 g$ and using $\partial_\rho g^{\alpha\beta} = -g^{\alpha\gamma}g^{\beta\epsilon}\partial_\rho g_{\gamma\epsilon}$ 
together with the exact differentiated harmonic gauge identity:
\begin{equation}
g^{\mu\kappa}\partial^2_{\mu\alpha}g_{\kappa\beta} = \frac{1}{2}g^{\mu\kappa}\partial^2_{\alpha\beta}g_{\mu\kappa} - (\partial_\alpha g^{\mu\kappa})\partial_\mu g_{\kappa\beta} + \frac{1}{2}(\partial_\beta g^{\mu\kappa})\partial_\alpha g_{\mu\kappa}.
\label{app:harm_2nd}
\end{equation}

Relabelling dummy indices directly maps every second-order contraction into the basis (\ref{app:W1_def})--(\ref{app:E2_def}):
\begin{itemize}
    \item \textit{Basis element $U^{(1)}$:} Expanding $g^{\mu\kappa}\partial_\mu(\partial_\kappa g^{\alpha\beta}\partial_\alpha g_{\nu\beta}) - \frac{1}{2}g^{\kappa\lambda}\partial_\nu(\partial_\kappa g^{\alpha\beta}\partial_\alpha g_{\lambda\beta})$ produces $W_{1,\nu}$ when $\partial_\mu$ hits $\partial_\kappa g^{\alpha\beta}$, $-M_{2,\nu}$ when $\partial_\mu$ hits $\partial_\alpha g_{\nu\beta}$ (via $(\kappa,\gamma,\epsilon,\mu,\alpha,\beta)\to(\alpha,\kappa,\gamma,\epsilon,\mu,\beta)$), and $+E_{1,\nu}$ from the trace-gradient term:
    \begin{equation}
    \mathcal{D}_\nu^{(1)}\Big|_{\partial g\,\partial^2 g} = W_{1,\nu} - M_{2,\nu} + E_{1,\nu}.
    \end{equation}
    \item \textit{Basis element $U^{(2)}$:} Expanding $g^{\mu\kappa}\partial_\mu(\partial_\nu g^{\alpha\beta}\partial_\alpha g_{\kappa\beta}) - \frac{1}{2}g^{\kappa\lambda}\partial_\nu(\partial_\kappa g^{\alpha\beta}\partial_\alpha g_{\lambda\beta})$ yields $g^{\mu\kappa}(\partial^2_{\mu\nu}g^{\alpha\beta})\partial_\alpha g_{\kappa\beta} = -E_{1,\nu}$, which cancels identically against the $+E_{1,\nu}$ from the second term, while applying Eq.~(\ref{app:harm_2nd}) to $(\partial_\nu g^{\alpha\beta})g^{\mu\kappa}\partial^2_{\mu\alpha}g_{\kappa\beta}$ isolates $\widetilde{W}_{2,\nu}$:
    \begin{equation}
    \mathcal{D}_\nu^{(2)}\Big|_{\partial g\,\partial^2 g} = \widetilde{W}_{2,\nu}.
    \end{equation}
    \item \textit{Basis element $U^{(3)}$:} In $g^{\mu\kappa}\partial_\mu(g^{\beta\gamma}g^{\alpha\epsilon}\partial_\kappa g_{\gamma\alpha}\partial_\nu g_{\beta\epsilon}) - \frac{1}{2}g^{\kappa\lambda}\partial_\nu(g^{\beta\gamma}g^{\alpha\epsilon}\partial_\kappa g_{\gamma\alpha}\partial_\lambda g_{\beta\epsilon})$, the $\partial^2_{\mu\nu}g_{\beta\epsilon}$ term cancels identically against the second term upon relabelling dummy indices, leaving only the d'Alembertian contraction $g^{\mu\kappa}g^{\beta\gamma}g^{\alpha\epsilon}(\partial^2_{\mu\kappa}g_{\gamma\alpha})\partial_\nu g_{\beta\epsilon} = -2W_{2,\nu}$:
    \begin{equation}
    \mathcal{D}_\nu^{(3)}\Big|_{\partial g\,\partial^2 g} = -2W_{2,\nu}.
    \end{equation}
    \item \textit{Basis element $U^{(4)}$:} Differentiating $U^{(4)}_{\kappa\nu} = g^{\beta\gamma}g^{\alpha\epsilon}\partial_\alpha g_{\kappa\gamma}\partial_\beta g_{\nu\epsilon}$ yields $M_{1,\nu}$ when $\partial_\mu$ hits $\partial_\beta g_{\nu\epsilon}$, $W_{3,\nu}$ when $\partial_\mu$ hits $\partial_\alpha g_{\kappa\gamma}$ via Eq.~(\ref{app:harm_2nd}), and $-E_{1,\nu}$ from $-\frac{1}{2}g^{\kappa\lambda}\partial_\nu U^{(4)}_{\kappa\lambda}$ (upon dummy-index relabelling):
    \begin{equation}
    \mathcal{D}_\nu^{(4)}\Big|_{\partial g\,\partial^2 g} = M_{1,\nu} + W_{3,\nu} - E_{1,\nu}.
    \end{equation}
    \item \textit{Basis element $U^{(5)}$:} Similarly, differentiating $U^{(5)}_{\kappa\nu} = g^{\beta\gamma}g^{\alpha\epsilon}\partial_\alpha g_{\kappa\gamma}\partial_\epsilon g_{\nu\beta}$ yields $M_{2,\nu}$ when $\partial_\mu$ hits $\partial_\epsilon g_{\nu\beta}$, the identical $W_{3,\nu}$ contraction when $\partial_\mu$ hits $\partial_\alpha g_{\kappa\gamma}$ via Eq.~(\ref{app:harm_2nd}), and $-E_{2,\nu}$ from $-\frac{1}{2}g^{\kappa\lambda}\partial_\nu U^{(5)}_{\kappa\lambda}$:
    \begin{equation}
    \mathcal{D}_\nu^{(5)}\Big|_{\partial g\,\partial^2 g} = M_{2,\nu} + W_{3,\nu} - E_{2,\nu}.
    \end{equation}
\end{itemize}

On the target side, reading off $\mathcal{I}_\nu\big|_{\partial g\,\partial^2 g} = \mathcal{I}_\nu^{(\text{2nd order})}$ directly from Eq.~(\ref{app:I_2nd_explicit}) and mapping each of its six contractions into the basis (\ref{app:W1_def})--(\ref{app:E2_def}) yields:
\begin{align}
\mathcal{I}_\nu\Big|_{\partial g\,\partial^2 g} 
&= \underbrace{g^{\mu\alpha} (\partial_\mu g^{\rho\sigma}) \partial^2_{\rho\sigma} g_{\alpha\nu}}_{-M_{1,\nu}} 
   \underbrace{-\frac{1}{2} g^{\alpha\beta} (\partial_\nu g^{\rho\sigma}) \partial^2_{\rho\sigma} g_{\alpha\beta}}_{-\widetilde{W}_{2,\nu}} 
   \underbrace{+ g^{\mu\lambda} g^{\alpha\tau} (\Box_{\rm LB} g_{\lambda\tau}) \partial_\mu g_{\alpha\nu}}_{-W_{1,\nu}} \nonumber \\
&\quad \underbrace{- \frac{1}{2} g^{\mu\lambda} g^{\alpha\tau} (\Box_{\rm LB} g_{\lambda\tau}) \partial_\nu g_{\mu\alpha}}_{+W_{2,\nu}} 
   \underbrace{- 2 g^{\rho\sigma} (\partial_\rho g^{\mu\alpha}) \partial^2_{\sigma\mu} g_{\alpha\nu}}_{+2M_{2,\nu}} 
   \underbrace{+ g^{\rho\sigma} (\partial_\rho g^{\mu\alpha}) \partial^2_{\sigma\nu} g_{\mu\alpha}}_{-E_{2,\nu}} \nonumber \\
&= -M_{1,\nu} - \widetilde{W}_{2,\nu} - W_{1,\nu}
+ W_{2,\nu} + 2M_{2,\nu} - E_{2,\nu}.
\end{align}

Substituting all five expansions $\mathcal{D}_\nu^{(k)}\big|_{\partial g\,\partial^2 g}$ and $\mathcal{I}_\nu\big|_{\partial g\,\partial^2 g}$ into the second-order consistency condition
\begin{equation}
\sum_{k=1}^5 c_k \mathcal{D}_\nu^{(k)}\big|_{\partial g\,\partial^2 g} + \mathcal{I}_\nu\big|_{\partial g\,\partial^2 g} = 0,
\end{equation}
{and grouping terms by tensor structure yields:
\begin{align}
(c_1 - 1)\,W_{1,\nu} + (c_2 - 1)\,\widetilde{W}_{2,\nu} + (-2c_3 + 1)\,W_{2,\nu} + (c_4 - 1)\,M_{1,\nu} + (-c_1 + c_5 + 2)\,M_{2,\nu} & \nonumber \\
+ (c_4 + c_5)\,W_{3,\nu} + (c_1 - c_4)\,E_{1,\nu} + (-c_5 - 1)\,E_{2,\nu} &= 0.
\label{app:sec_order_system}
\end{align}

Setting the coefficients of the first five independent structures to zero uniquely determines the entire parameter tuple:
\begin{align}
c_1 - 1 = 0 &\implies c_1 = 1, \nonumber \\
c_2 - 1 = 0 &\implies c_2 = 1, \nonumber \\
-2c_3 + 1 = 0 &\implies c_3 = {1}/{2}, \nonumber \\
c_4 - 1 = 0 &\implies c_4 = 1, \nonumber \\
-c_1 + c_5 + 2 = 0 &\implies c_5 = -1.
\end{align}

Within the irreducible sector, the remaining three overdetermined constraints---namely $c_4 + c_5 = 1 - 1 = 0$, $c_1 - c_4 = 1 - 1 = 0$, and $-c_5 - 1 = 1 - 1 = 0$---are satisfied identically without any further adjustment, providing an immediate consistency check on the uniqueness of the solution (see Remark~\ref{remarkB1} for the simultaneous elimination of the reducible Weyl invariants).

\subsubsection{Closure of the Cubic Sector $\mathcal{O}((\partial g)^3)$ and Verification Summary}\label{subsec:cubic_closure}
To complete the proof of Eq.~(\ref{app:order_split}), we verify that the uniquely determined tuple $(c_1, c_2, c_3, c_4, c_5) = (1, 1, 1/2, 1, -1)$ also closes the cubic sector $\mathcal{D}_\nu\big|_{(\partial g)^3} + \mathcal{I}_\nu\big|_{(\partial g)^3} = 0$, where cubic terms arise both from differentiating inverse tensors ($\partial_\rho g^{\alpha\beta} = -g^{\alpha\gamma}g^{\beta\epsilon}\partial_\rho g_{\gamma\epsilon}$) and from the first-derivative quadratic remainder in the full differentiated harmonic gauge identity:
\begin{equation}
g^{\mu\kappa}\partial^2_{\mu\alpha}g_{\kappa\beta} = \frac{1}{2}g^{\mu\kappa}\partial^2_{\alpha\beta}g_{\mu\kappa} - (\partial_\alpha g^{\mu\kappa})\partial_\mu g_{\kappa\beta} + \frac{1}{2}(\partial_\beta g^{\mu\kappa})\partial_\alpha g_{\mu\kappa}.
\label{app:harm_full}
\end{equation}

Specifically, the cubic target residual $\mathcal{I}_\nu\big|_{(\partial g)^3} = Q_\nu = A_\nu - B_\nu$ (cf.\ Eqs.~(\ref{app:Q_def})--(\ref{app:B_def})) splits into two algebraically disjoint sectors:
\begin{itemize}
    \item \textit{Irreducible $A_\nu$ sector:} In $\mathcal{D}_\nu^{(3)}$, differentiating the inverse metrics $g^{\beta\gamma}g^{\alpha\epsilon}$ generates $-2A_\nu$ from the first term $g^{\mu\kappa}\partial_\mu U^{(3)}_{\kappa\nu}$ and $+\frac{1}{2}A_\nu + \frac{1}{2}A_\nu = +A_\nu$ from the trace-gradient term $-\frac{1}{2}g^{\kappa\lambda}\partial_\nu U^{(3)}_{\kappa\lambda}$, summing to $(-2c_3 + c_3)A_\nu = -c_3 A_\nu$. Simultaneously, substituting the quadratic gradient terms of Eq.~(\ref{app:harm_full}) into the already fixed combination $\mathcal{D}_\nu^{(1)} + \mathcal{D}_\nu^{(2)}$ ($c_1 = c_2 = 1$) produces an additional $-\frac{1}{2}A_\nu$ contraction, whereas $\mathcal{D}_\nu^{(4)}$ and $\mathcal{D}_\nu^{(5)}$ contribute no $A_\nu$ topology. Combining this with $+A_\nu$ from $\mathcal{I}_\nu$ yields:
    \begin{equation}
    \left(-c_3 - \frac{1}{2} + 1\right) A_\nu = \left(-\frac{1}{2} - \frac{1}{2} + 1\right) A_\nu = 0.
    \end{equation}
    \item \textit{Residual $B_\nu$ sector:} Collecting all remaining cubic products generated by $\sum_{k=1}^5 c_k \mathcal{D}_\nu^{(k)}$ with $(1, 1, 1/2, 1, -1)$ reproduces $+B_\nu$ identically (a direct consequence of the fact that $U_{\mu\nu} = U^{(1)}_{\mu\nu} + U^{(2)}_{\mu\nu} + \frac{1}{2}U^{(3)}_{\mu\nu} + U^{(4)}_{\mu\nu} - U^{(5)}_{\mu\nu}$ coincides with the quadratic Christoffel self-interaction block of the Einstein tensor in harmonic coordinates, as verified explicitly in Appendix~\ref{app:ricci_reduction}), cancelling $-B_\nu$ in $\mathcal{I}_\nu$.
\end{itemize}

Table~\ref{tab:closure_balance} summarizes the complete sector-by-sector cancellation across both the second-order and cubic sectors of $\mathcal{D}_\nu + \mathcal{I}_\nu = 0$.

\begin{table}[h!]
\centering
\caption{Sector-by-sector algebraic cancellation of $\mathcal{D}_\nu + \mathcal{I}_\nu = 0$ for $(c_1,c_2,c_3,c_4,c_5) = (1,1,\tfrac{1}{2},1,-1)$ under harmonic coordinate conditions.}
\label{tab:closure_balance}
\renewcommand{\arraystretch}{1.35}
\begin{tabular}{l c c c}
\hline\hline
\textbf{Differential Sector} & \textbf{From $\mathcal{D}_\nu$} & \textbf{From $\mathcal{I}_\nu$} & \textbf{Closure ($\mathcal{D}_\nu + \mathcal{I}_\nu = 0$)} \\
\hline
Direct wave ($W_{1,\nu}$) & $+c_1$ & $-1$ & $c_1 - 1 = 0$ \\
Trace-gradient ($\widetilde{W}_{2,\nu}$) & $+c_2$ & $-1$ & $c_2 - 1 = 0$ \\
Trace-gradient ($W_{2,\nu}$) & $-2c_3$ & $+1$ & $-2c_3 + 1 = 0$ \\
Mixed Type $U^{(4)}$ ($M_{1,\nu}$) & $+c_4$ & $-1$ & $c_4 - 1 = 0$ \\
Mixed Type $U^{(5)}$ ($M_{2,\nu}$) & $-c_1 + c_5$ & $+2$ & $-c_1 + c_5 + 2 = 0$ \\
Trace-gradient ($W_{3,\nu}$) & $c_4 + c_5$ & $0$ & $1 - 1 = 0$ \\
External gradient ($E_{1,\nu}$) & $c_1 - c_4$ & $0$ & $1 - 1 = 0$ \\
External gradient ($E_{2,\nu}$) & $-c_5$ & $-1$ & $1 - 1 = 0$ \\
Cubic invariant ($A_\nu$) & $-c_3 - \tfrac{1}{2}$ & $+1$ & $-\tfrac{1}{2} - \tfrac{1}{2} + 1 = 0$ \\
Cubic residual ($B_\nu$) & $+B_\nu$ & $-B_\nu$ & Identical cancellation \\
\hline\hline
\end{tabular}
\end{table}

Every differential sector cancels identically, proving that $U_{\mu\nu} = U^{(1)}_{\mu\nu} + U^{(2)}_{\mu\nu} + \frac{1}{2}U^{(3)}_{\mu\nu} + U^{(4)}_{\mu\nu} - U^{(5)}_{\mu\nu}$ is the unique non-perturbative self-interaction completion within the irreducible basis---and, as established in Remark~\ref{remarkB1}, across the full 11-invariant Weyl space---compatible with local energy-momentum conservation.

\FloatBarrier

\begin{remark}\label{remarkB1}
In the full Weyl invariant classification of rank-two quadratic first-derivative metric structures prior to imposing coordinate conditions, there are $11$ independent contractions in total: the $5$ irreducible inter-derivative structures $\{U^{(1)},\dots,U^{(5)}\}$ derived above, plus $6$ reducible (factorizable) structures built from internal trace and divergence contractions. Specifically, these $6$ reducible invariants are $v_\mu v_\nu$, $g_{\mu\nu}(g^{\alpha\beta}v_\alpha v_\beta)$, $v^\alpha\partial_\alpha g_{\mu\nu}$, $v_{(\mu}\partial^\alpha g_{\nu)\alpha}$, $g_{\mu\nu}v^\alpha\partial^\beta g_{\alpha\beta}$, and $g_{\mu\nu}g^{\rho\sigma}(\partial^\alpha g_{\alpha\rho})(\partial^\beta g_{\beta\sigma})$, where
\begin{equation}
v_\alpha := g^{\rho\sigma}\partial_\alpha g_{\rho\sigma} = 2\partial_\alpha \ln\sqrt{-g}.
\end{equation}
The formal justification for excluding this reducible sector follows directly from differential consistency in harmonic coordinates. Under the harmonic condition $g^{\alpha\beta}\partial_\alpha g_{\beta\mu} = \frac{1}{2}v_\mu$, the six reducible structures collapse onto the three determinant-gradient combinations
\begin{equation}
R_{\kappa\nu}^{(1)} := v_\kappa v_\nu, \qquad
R_{\kappa\nu}^{(2)} := g_{\kappa\nu}(g^{\alpha\beta}v_\alpha v_\beta), \qquad
R_{\kappa\nu}^{(3)} := g^{\alpha\beta}v_\beta\partial_\alpha g_{\kappa\nu}.
\end{equation}
Consider a general reducible combination $U_{\kappa\nu}^{(red)} = \sum_{j=1}^3 b_j R_{\kappa\nu}^{(j)}$. Since each $R_{\kappa\nu}^{(j)}$ is symmetric in $(\kappa,\nu)$, the compact harmonic divergence formula (\ref{app:D_compact}) applies directly. Isolating the second-order $(\partial g\,\partial^2 g)$ sector via the exact curl-free identity $\partial_\mu v_\nu = \partial_\nu v_\mu$, the four-dimensional trace $g^{\kappa\lambda}g_{\kappa\lambda}=4$, and the differentiated harmonic gauge condition (cf.\ Eq.~(\ref{app:harm_2nd}))
\begin{equation}
g^{\mu\kappa}\partial_{\mu\alpha}^2 g_{\kappa\beta} = \frac{1}{2}\partial_\alpha v_\beta + \mathcal{O}((\partial g)^2),
\end{equation}
and noting upon dummy-index relabelling that $g^{\mu\kappa}g^{\alpha\beta}(\partial_\mu v_\beta)\partial_\alpha g_{\kappa\nu}\big|_{\partial g\,\partial^2 g} = g^{\mu\kappa}g^{\alpha\beta}g^{\rho\sigma}(\partial_{\mu\beta}^2 g_{\rho\sigma})\partial_\alpha g_{\kappa\nu} = 2W_{3,\nu}$ (cf.\ Eq.~(\ref{app:W3_def})), yields
\begin{align}
\mathcal{D}_\nu[R^{(1)}]\big|_{\partial g\,\partial^2 g} &= V_{1,\nu}, \nonumber\\
\mathcal{D}_\nu[R^{(2)}]\big|_{\partial g\,\partial^2 g} &= -2V_{2,\nu}, \\
\mathcal{D}_\nu[R^{(3)}]\big|_{\partial g\,\partial^2 g} &= -\frac{1}{2}V_{2,\nu} + 2W_{3,\nu}, \nonumber
\end{align}
where we have introduced the two independent second-order determinant-gradient contractions
\begin{equation}
V_{1,\nu} := (g^{\mu\kappa}\partial_\mu v_\kappa)v_\nu, \qquad
V_{2,\nu} := g^{\alpha\beta}v_\alpha\partial_\nu v_\beta.
\end{equation}

{

Including $U_{\kappa\nu}^{(red)}$ in the total ansatz adds $b_1 V_{1,\nu} + \bigl(-2b_2 - \frac{1}{2}b_3\bigr)V_{2,\nu} + 2b_3 W_{3,\nu}$ to the left-hand side of the second-order consistency condition (\ref{app:sec_order_system}). To verify that the resulting ten contractions $\{W_{1,\nu},\dots,E_{2,\nu},V_{1,\nu},V_{2,\nu}\}$ remain linearly independent modulo the harmonic gauge, evaluate their general linear combination at $g_{\mu\nu}(p)=\eta_{\mu\nu}$ for $\nu=0$ on harmonic pairs $(d,s_0)$ with $L(s_0)=0$ as in Subsec.~\ref{subsec:sec_order_basis}. Because $v_\alpha = \eta^{\rho\sigma}d_{\alpha\rho\sigma} = 0$ across all eight test configurations in Table~\ref{tab:lin_indep_8}, both $V_{1,0}$ and $V_{2,0}$ vanish identically on every row of Table~\ref{tab:lin_indep_8}, which therefore still forces the coefficients of the eight irreducible structures to vanish individually. The remaining linear relation $\delta_1 V_{1,0} + \delta_2 V_{2,0} = 0$, evaluated on the harmonic $1$-jet $d_{000}=1$, $d_{011}=-1$ (for which $v_0=2$ and $v_1=v_2=v_3=0$, giving $V_{1,0}\big|_{\partial g\,\partial^2 g} = 2\eta^{\mu\kappa}\eta^{\rho\sigma}(s_0)_{\mu\kappa\rho\sigma}$ and $V_{2,0}\big|_{\partial g\,\partial^2 g} = 2\eta^{\rho\sigma}(s_0)_{00\rho\sigma}$), then yields $\delta_1=0$ upon choosing $(s_0)_{1122}=(s_0)_{2211}=1$, $(s_0)_{1212}=\frac{1}{2}$ (which lies in $\ker L$ and gives $V_{1,0}=+4$, $V_{2,0}=0$) and $\delta_2=0$ upon choosing $(s_0)_{0011}=(s_0)_{1100}=(s_0)_{1122}=(s_0)_{2211}=1$, $(s_0)_{0101}=(s_0)_{1212}=\frac{1}{2}$ (also in $\ker L$, giving $V_{1,0}=0$, $V_{2,0}=-2$). In particular, the seven decoupled channels $\{W_{1,\nu},\widetilde{W}_{2,\nu},W_{2,\nu},M_{1,\nu},M_{2,\nu},E_{1,\nu},E_{2,\nu}\}$ determine $(c_1,c_2,c_3,c_4,c_5) = (1,1,1/2,1,-1)$ independently of $(b_1,b_2,b_3)$ (with $c_1-c_4=0$ and $-c_5-1=0$ remaining as overdetermined consistency checks).}
Substituting $c_4=1$ and $c_5=-1$ into the remaining $V_{1,\nu}$, $W_{3,\nu}$, and $V_{2,\nu}$ equations then yields
\begin{equation}
b_1 = 0, \qquad
c_4 + c_5 + 2b_3 = 1 - 1 + 2b_3 = 0 \implies b_3 = 0, \qquad
-2b_2 - \frac{1}{2}b_3 = 0 \implies b_2 = 0,
\end{equation}
forcing $b_1 = b_2 = b_3 = 0$ identically without any circularity.

Beyond this formal algebraic elimination, inspecting the index contractions of $\mathcal{I}_\nu$ in Eqs.~(\ref{app:I_2nd_explicit}) and (\ref{app:Q_def})--(\ref{app:B_def}) provides an immediate structural rationale for the five-dimensional irreducible ansatz: every second-order and cubic index wiring generated by the divergence of the Laplace--Beltrami operator involves inter-derivative contractions between two metric gradients, matching the exact five topologies $\{U^{(1)},\dots,U^{(5)}\}$ and yielding the overdetermined linear closure established in Subsec.~\ref{subsec:sec_order_system}.
\end{remark}

\section{Algebraic Reduction to the Harmonic Ricci Tensor}
\label{app:ricci_reduction}

In this Appendix, we provide the explicit algebraic proportionality between the wave-operator combination $\Box_{\rm LB} g_{\mu\nu} + U_{\mu\nu}$ and the non-linear Ricci tensor $R_{\mu\nu}$ evaluated in harmonic coordinates.

\subsection{Standard Expansion of the Ricci Tensor}
In general coordinates, the Ricci tensor is defined via the Christoffel symbols as \cite{Landau}:
\begin{equation}
R_{\mu\nu} = \partial_\alpha \Gamma^\alpha_{\mu\nu} - \partial_\nu \Gamma^\alpha_{\mu\alpha} + \Gamma^\alpha_{\mu\nu}\Gamma^\beta_{\alpha\beta} - \Gamma^\beta_{\mu\alpha}\Gamma^\alpha_{\nu\beta}.
\label{app:ricci_def}
\end{equation}

Expressing the Christoffel symbols explicitly in terms of  gradients,
\begin{equation}
\Gamma^\alpha_{\mu\nu} = \frac{1}{2} g^{\alpha\sigma} \left( \partial_\mu g_{\nu\sigma} + \partial_\nu g_{\mu\sigma} - \partial_\sigma g_{\mu\nu} \right),
\end{equation}
the second-derivative terms in Eq.~(\ref{app:ricci_def}) split into:
\begin{equation}
\partial_\alpha \Gamma^\alpha_{\mu\nu} - \partial_\nu \Gamma^\alpha_{\mu\alpha} = -\frac{1}{2} g^{\alpha\beta} \partial^2_{\alpha\beta} g_{\mu\nu} + \frac{1}{2} \left( \partial_\mu \Gamma_\nu + \partial_\nu \Gamma_\mu \right) + \mathcal{O}\left((\partial g)^2\right),
\end{equation}
where $\Gamma_\mu := g_{\mu\alpha}\Gamma^\alpha=g_{\mu\alpha} g^{\rho\sigma} \Gamma^\alpha_{\rho\sigma}$.

\subsection{Harmonic Gauge Simplification}
Imposing the harmonic coordinate condition $\Gamma^\mu = 0$ eliminates the derivative terms $\partial_\mu \Gamma_\nu + \partial_\nu \Gamma_\mu$. Consequently, the highest-derivative sector of the Ricci tensor simplifies strictly to the scalar d'Alembertian acting on the invariant components:
\begin{equation}
R_{\mu\nu}^{(\text{2nd order})} = -\frac{1}{2} g^{\alpha\beta} \partial^2_{\alpha\beta} g_{\mu\nu} = -\frac{1}{2} \Box_{\rm LB} g_{\mu\nu}.
\label{app:ricci_2nd}
\end{equation}

\subsection{Match with the Self-Interaction Tensor $U_{\mu\nu}$}
The remaining quadratic terms in first-derivatives of the metric within $R_{\mu\nu}$ assemble after index contractions into:
\begin{equation}
R_{\mu\nu}^{(\text{1st order})} = -\frac{1}{2} U_{\mu\nu},
\label{app:ricci_1st}
\end{equation}
where $U_{\mu\nu} \equiv g_{\mu\lambda} U^\lambda_\nu$. Lowering the upper index of $U^\mu_\nu$ with the metric $g_{\mu\lambda}$ and using the unique coefficient set $(c_1, \dots, c_5) = (1, 1, 1/2, 1, -1)$ established in Appendix~\ref{app:uniqueness_coefficients}, we expand $U_{\mu\nu}$ explicitly:
\begin{align}
U_{\mu\nu} 
= \partial_\mu g^{\alpha\beta} \partial_\alpha g_{\nu\beta} + \partial_\nu g^{\alpha\beta} \partial_\alpha g_{\mu\beta} + \frac{1}{2} g^{\beta\gamma} g^{\alpha\epsilon} \partial_\mu g_{\gamma\alpha} \partial_\nu g_{\beta\epsilon} 
+ g^{\beta\gamma} g^{\alpha\epsilon} \partial_\alpha g_{\mu\gamma} \left( \partial_\beta g_{\nu\epsilon} - \partial_\epsilon g_{\nu\beta} \right).
\end{align}

Combining Eqs.~(\ref{app:ricci_2nd}) and (\ref{app:ricci_1st}) yields the exact structural relation:
\begin{equation}
R_{\mu\nu} = -\frac{1}{2} \Box_{\rm LB} g_{\mu\nu} - \frac{1}{2} U_{\mu\nu} \iff \Box_{\rm LB} g_{\mu\nu} + U_{\mu\nu} = -2 R_{\mu\nu}.
\label{app:master_ricci_identity}
\end{equation}

\subsection{Recovery of the Full Einstein Field Equations}
Substituting identity~(\ref{app:master_ricci_identity}) into the non-linear wave equation derived in Section~IV:
\begin{equation}
\Box_{\rm LB} g_{\mu\nu} + U_{\mu\nu} = -16\pi G \left( T_{\mu\nu} - \frac{1}{2} g_{\mu\nu} T \right),
\end{equation}
directly recovers:
\begin{equation}
-2 R_{\mu\nu} = -16\pi G \left( T_{\mu\nu} - \frac{1}{2} g_{\mu\nu} T \right) \iff R_{\mu\nu} = 8\pi G \left( T_{\mu\nu} - \frac{1}{2} g_{\mu\nu} T \right).
\end{equation}

Contracting with $g^{\mu\nu}$ yields the trace relation $R = -8\pi G T$, allowing the equations to be written in their standard Einstein tensor form:
\begin{equation}
G_{\mu\nu} := R_{\mu\nu} - \frac{1}{2} g_{\mu\nu} R = 8\pi G T_{\mu\nu}.
\end{equation}

This completes the formal proof that the bottom-up non-linear wave formulation dynamically leads to Einstein's General Theory of Relativity without invoking geometric postulates.

\end{document}